\documentclass[a4paper,12pt]{article}
\usepackage{atlasphysics}

\usepackage[displaymath]{lineno}
\RequirePackage{graphicx}
\usepackage{jheppub}

\usepackage{lineno}
\usepackage{xspace}
\usepackage{cancel}
\usepackage{slashbox}
\usepackage{multirow}
\usepackage{array}
\usepackage{comment}
\usepackage{mathtools}
\usepackage{amsmath}

\usepackage[tight]{subfigure}

\newcommand{\sbottomTitle}{Search for direct third-generation squark pair production in final states with missing transverse momentum and two $\boldsymbol{b}$-jets in $\boldsymbol{\sqrt{s} = 8}$ TeV $\boldsymbol{pp}$ collisions with the ATLAS detector.}

\def\SRMCT{SRA\xspace}
\def\SRISR{SRB\xspace}
\def\CRAL{CRA\_1L\xspace}
\def\CRBL{CRB\_1L\xspace}
\def\CRASF{CRA\_SF\xspace}
\def\CRBSF{CRB\_SF\xspace}
\def\CRADF{CRA\_DF\xspace}

\newcommand{\bone}   {\ensuremath{\tilde{b}^{}_{1}}\xspace}
\newcommand{\tone}   {\ensuremath{\tilde{t}^{}_{1}}\xspace}
\newcommand{\neut}  {\ensuremath{\tilde{\chi}^{0}_{1}}\xspace}  
\newcommand{\chipm}  {\ensuremath{\tilde{\chi}^{\pm}_{1}}\xspace} 
\def\etmiss{\ensuremath{E_{\mathrm{T}}^{\mathrm{miss}}}\xspace}

\def\dphimin{\ensuremath{\mathrm{\Delta}\phi_{\mathrm{min}}}}
\def\trmass{\ensuremath{\mathrm{m}_\mathrm{T}}}
\def\bjets{$b$-jets\xspace}

\def\btagging{$b$-tagging\xspace}

\def\btagged{$b$-tagged\xspace}
\def\bhadron{$b$-hadron\xspace}

\def\bquark{$b$-quark\xspace}
\def\bquarks{$b$-quarks\xspace}
\def\meff{\ensuremath{m_{\mathrm{eff}}}\xspace}
\def\mt{\ensuremath{m_{\mathrm{T}}}\xspace}

\def\HT3{\ensuremath{H_{{\rm T},3}}\xspace}
\def\emf{\ensuremath{f_{{\rm em}}}\xspace}
\def\chf{\ensuremath{f_{{\rm ch}}}\xspace}

\newcommand{\gl}   {\tilde{g}}
\newcommand{\sq}   {\tilde{q}}
\newcommand{\ttW}{\ensuremath{t\bar{t}+W}\xspace}
\newcommand{\ttZ}{\ensuremath{t\bar{t}+Z}\xspace}
\def\Wjets{$W$+jets\xspace}
\def\Zjets{$Z$+jets\xspace}

\def\WHF{$W$+hf\xspace}
\def\ZHF{$Z$+hf\xspace}
\renewcommand{\ttbar} {\ensuremath{t\bar{t}}\xspace}

\renewcommand{\met} {\ensuremath{E_{\mathrm{T}}^{\mathrm{miss}}}\xspace}
\renewcommand{\pt} {\ensuremath{p_\mathrm{T}}\xspace}
\newcommand{\mct}{\ensuremath{m_\mathrm{CT}}\xspace}
\newcommand{\mbb}{\ensuremath{m_{bb}}\xspace}
\def\mll{\ensuremath{m_{\ell \ell}}\xspace}
\newcommand{\lumi}{20.1}
\newcommand\totallumi{20.1~${\rm fb}^{-1}$}
\newcommand\lowptthr{30}
\newcommand\pythia{{\sc Pythia}\xspace}

\newcommand\herwigjimmy{{\sc Herwig}+{\sc Jimmy}\xspace}
\newcommand\sherpa{{\sc Sherpa}\xspace}
\newcommand\powheg{{\sc Powheg}\xspace}
\newcommand\alpgen{{\sc Alpgen}\xspace}

\def\met{\ensuremath{E_{\mathrm{T}}^{\mathrm{miss}}}\xspace}
\def\ptmissvec{\ensuremath{\textit{{\textbf{p}}}_{\mathrm{T}}^{\mathrm{miss}}}\xspace}
\def\CLs{\ensuremath{CLs}\xspace}

\title{\sbottomTitle}
\author{The ATLAS Collaboration}

\def\FinalAbstract{The results of a search for
pair production of supersymmetric partners of the Standard Model third-generation quarks are reported. This search uses \totallumi~of $pp$ collisions at $\sqrt{s}=8$~TeV collected by the ATLAS experiment at the Large Hadron Collider.
The lightest bottom  and top squarks (\bone\ and \tone\ respectively) are searched for in a final state with large
missing transverse momentum and two jets identified as originating from $b$-quarks.
No excess of events above the expected level of Standard Model background is found. The results are used to set upper limits on the visible cross section for processes beyond the Standard Model.
Exclusion limits at the 95\% confidence level on the masses of the third-generation squarks are derived in phenomenological supersymmetric $R$-parity-conserving models in which 
either the bottom or the top squark is the lightest squark. The $\bone$ is assumed to decay via $\bone \rightarrow b \neut$ and the $\tone$ via $\tone \rightarrow b \chipm$, with undetectable products of the subsequent decay of the \chipm due to the small mass splitting between the \chipm and the \neut.}

\abstract{\FinalAbstract}

\usepackage{preprintcover}  
\PreprintCoverPaperTitle{\sbottomTitle}  
\PreprintIdNumber{CERN-PH-EP-2013-119}  
\PreprintCoverAbstract{\FinalAbstract}  
\PreprintJournalName{JHEP}  

\begin{document}

\maketitle

\section{Introduction}
Supersymmetry (SUSY)~\cite{Miyazawa:1966,Ramond:1971gb,Golfand:1971iw,Neveu:1971rx,Neveu:1971iv,Gervais:1971ji,Volkov:1973ix,Wess:1973kz,Wess:1974tw} 
 provides an extension of the Standard Model (SM) that 
solves the hierarchy problem \cite{Weinberg:1975gm,Gildener:1976ai,Weinberg:1979bn,Susskind:1978ms} by 
introducing supersymmetric partners of  the known 
bosons and fermions. In the framework of the $R$-parity-conserving minimal supersymmetric extension of the SM (MSSM)~\cite{Fayet:1976et,Fayet:1977yc,Farrar:1978xj,Fayet:1979sa,Dimopoulos:1981zb}, 
SUSY particles are produced in pairs and the lightest supersymmetric particle (LSP) is stable, providing a possible candidate 
for dark matter. In a large variety of models, the LSP is the lightest neutralino ($\neut$).
The coloured superpartners of quarks and gluons, the squarks ($\sq$) and the gluinos ($\gl$), if not too heavy,  would be produced in strong interaction processes at the Large Hadron Collider (LHC)~\cite{Evans:2008zzb} and decay via cascades ending with the LSP.
The undetected LSP would result in missing transverse momentum  while the rest of the cascade would yield final states with  multiple jets and possibly leptons.

A study of the expected SUSY particle spectrum derived from naturalness considerations~\cite{Barbieri:1987fn,deCarlos:1993yy} suggests that the supersymmetric partners of the third-generation SM quarks are the lightest coloured supersymmetric particles. This may lead to the lightest bottom squark (sbottom, $\bone$) and 
top squark (stop, $\tone$) mass eigenstates being significantly lighter than the other squarks and the gluinos. As a consequence, $\bone$ and $\tone$ could be pair-produced with relatively large cross sections at the LHC. 

Two possible sets of SUSY mass spectra are considered in this paper.
In the first set of scenarios, the lightest sbottom is the only coloured sparticle contributing to the production processes and it only decays via $\bone \rightarrow b\neut$. In the second set, the lightest stop is the only coloured sparticle allowed in the production processes and it decays exclusively via $\tilde{t}_1\to b \chipm$,
where the lightest chargino (\chipm) decays via a virtual $W$ boson into the three-body final state $\neut f\bar{f}'$. 
In the cases considered in this article, the fermions
$f$ and $f'$  may have
transverse momenta below the reconstruction thresholds applied in the analysis, as a consequence of a  small value for  
$\Delta m\equiv m_{\chipm}-m_{\neut}$.

In both scenarios, events are characterised by the presence of two jets originating from the hadronisation of the \bquarks
and large missing transverse momentum.
Results of searches for direct sbottom and stop production have been previously reported by the 
ATLAS~\cite{PRL_bjet,Aad:2012uu,Aad:2012yr,Aad:2012tx,Aad:2012xqa,Aad:2012ywa} 
and CMS~\cite{Chatrchyan:2012paa,Chatrchyan:2012wa,Chatrchyan:2013lya} experiments at the LHC, and by the 
Tevatron~\cite{Aaltonen:2010dy,Abazov:2010wq} and 
LEP~\cite{lep2} experiments.

\section{The ATLAS detector and data samples}
The ATLAS detector~\cite{DetectorPaper:2008} consists of inner tracking devices
surrounded by a superconducting solenoid, electromagnetic
and hadronic calorimeters and a muon spectrometer
with a magnetic field produced by a set of toroids. The inner detector (ID),
in combination with a superconducting solenoid magnet with a central field of 2~T,
provides precision tracking and momentum measurement of charged particles in a pseudorapidity\footnote{ATLAS uses a right-handed coordinate 
system with its origin at the nominal interaction point (IP) in the centre of the detector and the $z$-axis along 
the beam pipe. The $x$-axis points from the IP to the centre of the LHC ring, and the $y$-axis points upward. 
Cylindrical coordinates $(r,\phi)$ are used in the transverse plane, $\phi$ being the azimuthal angle around the 
beam pipe. The pseudorapidity is defined in terms of the polar angle $\theta$ as $\eta=-\ln\tan(\theta/2)$. The 
distance~$\mathrm{\Delta} R$ in the $\eta$--$\phi$ space is defined as 
$\mathrm{\Delta} R = \sqrt{(\mathrm{\Delta}\eta)^2+(\mathrm{\Delta}\phi)^2}$.} range $|\eta| < 2.5$ and allows efficient 
identification of jets originating from  \bhadron decays using
impact parameter measurements and reconstructed secondary
decay vertices. The ID consists of a silicon pixel detector, a silicon microstrip
detector and a straw tube tracker ($|\eta| < 2.0$) that also provides transition
radiation measurements for electron identification.
The calorimeter system covers the pseudorapidity range
$|\eta| < 4.9$. It is composed of sampling calorimeters with
either liquid argon or scintillating tiles as the active
medium. The muon spectrometer has separate trigger and high-precision 
tracking chambers, the latter providing muon identification and momentum measurement for $|\eta| < 2.7$.

The data sample used in this analysis was taken during the period from March to December 2012 with the LHC operating at a $pp$ centre-of-mass energy of $\sqrt{s}=8$~TeV. 
Candidate signal events are selected using a trigger based on a missing transverse momentum selection (\met), which is found to be 99\% 
efficient
for events passing the offline selection of  $\met> 150$~\GeV.
The trigger efficiency variations over data-taking periods are measured to be less than 1\% after the offline requirements. 
After the application of beam, detector, and data-quality requirements, the total integrated luminosity considered is \lumi~\ifb. The uncertainty on the integrated luminosity is $\pm 2.8 $\%. It is derived, following the same methodology as that detailed in ref.~\cite{Aad:2013ucp}, from a preliminary 
calibration of the luminosity scale using beam-separation scans performed in November 2012. 
Events with final-state electrons or muons that satisfy single-lepton or dilepton triggers are used to define control regions in a total data sample of 20.3~\ifb. 
A requirement on the transverse momentum $\pt>25$~GeV  is applied to the highest-\pt electron or 
muon to ensure the trigger selection is fully efficient.

\section{Simulated event samples}
\label{sec:mc}
Simulated Monte Carlo (MC) event samples are used to aid in the
description of the background and to model the SUSY signal. All SM MC
samples utilised in the analysis are produced using the ATLAS
Underlying Event Tune 2B~\cite{mc11ctunes} and are processed through the ATLAS detector simulation~\cite{atlassimulation}
based on GEANT4~\cite{geant4} or passed through a fast simulation using a parameterisation of the performance of the ATLAS electromagnetic and hadronic calorimeters~\cite{ATLAS:atlfast}. 
The effect of multiple $pp$ interactions per bunch crossing (pile-up) is also taken into account.

The top-quark pair (\ttbar) background is simulated with {\sc Powheg-1.0}~\cite{powheg}  interfaced to
{\sc Pythia-6.426}~\cite{pythia} for the fragmentation and hadronisation processes. The top-quark mass is fixed
at 172.5 GeV, and the next-to-leading-order (NLO) parton distribution
function (PDF) set CT10~\cite{Nadolsky:2008zw} is used. Samples from {\sc Alpgen-2.14}~\cite{alpgen} and \powheg each interfaced to {\sc Herwig-6.520} 
 for the fragmentation and hadronisation processes, including {\sc Jimmy-4.31}~\cite{JIMMY} for the underlying event 
description,  are used to estimate the 
generator and fragmentation systematic uncertainties, while {\sc AcerMC-3.8}~\cite{Kersevan:2013ji} interfaced to \pythia samples are
used to estimate the showering uncertainties.
Single top-quark production for the $s$-channel and $Wt$ processes is simulated with {\sc Mc@Nlo-4.06} interfaced to \herwigjimmy, while the $t$-channel process is simulated with {\sc AcerMC} interfaced to \pythia and using the CTEQ6L1~\cite{Nadolsky:2008zw} PDF set.
Samples of $\ttbar$+$W$/$Z$ events are generated using the leading-order generator {\sc MadGraph-5.1.4.8}~\cite{MadGraph} interfaced to 
\pythia for the fragmentation and hadronisation processes.
Samples of $Z/\gamma^*$ or $W$ production, both in association with up to five jets, are produced with {\sc Sherpa-1.4.1}~\cite{Gleisberg:2008ta}.  
MC samples of dibosons ($ZZ$, $WZ$ and $WW$) are generated using \sherpa.

The background predictions are normalised to theoretical cross sections, calculated including higher-order QCD corrections where available,
and are compared to data in appropriate control regions. The inclusive cross sections for \Zjets and \Wjets processes are calculated with {\sc DYNNLO}~\cite{Catani:2009sm} with the MSTW 2008 next-to-next-to-leading-order PDF set~\cite{Martin:2009iq}. 
Approximate NLO+NNLL (next-to-next-to-leading-logarithm) cross sections are used in the normalisation of the \ttbar~\cite{Aliev:2010zk} and $Wt$~\cite{Kidonakis:2010ux} samples. Cross sections calculated at NLO are used for the \ttW and \ttZ samples~\cite{Garzelli:2012bn} and for the diboson samples~\cite{Binoth:2006mf}.

The SUSY signal samples are generated using 
{\sc MadGraph} interfaced to \pythia (with PDF set
{\sc CTEQ6L1})  to ensure an accurate treatment of the initial-state radiation (ISR).
Additional samples with different ISR parameter values are generated to evaluate the ISR systematic uncertainty. Polarisation effects due to the choice of left- and right-handed scalar sbottom or stop mixing were found to have a negligible impact on the analysis.
Signal cross sections are calculated to next-to-leading-order in the strong coupling constant, 
adding the resummation of soft gluon emission at next-to-leading-logarithmic 
accuracy (NLO+NLL)~\cite{Beenakker:1997ut,Beenakker:2010nq,Beenakker:2011fu}. 
The nominal cross section and its uncertainty are taken from an envelope of cross-section 
predictions using different PDF sets and factorisation and renormalisation scales, as described in ref.~\cite{Kramer:2012bx}.

\section{Physics object reconstruction}
Jets are reconstructed from three-dimensional cell-energy clusters in the calorimeter 
 using the anti-$k_t$ jet algorithm~\cite{Cacciari:2008gp,Cacciari:2005hq} 
with a radius parameter of 0.4. 
The measured jet energy is corrected for inhomogeneities and for the non-compensating nature of the calorimeter by weighting  
energy deposits arising from electromagnetic and hadronic showers by correction factors derived from 
MC simulations and validated with data. 
 An additional calibration is subsequently applied to the corrected jet energies, relating the
response of the calorimeter to the true jet energy~\cite{Aad:2011he}.
 The impact of additional collisions in the same or neighbouring bunch crossings is accounted for using corrections derived as a function of the average number of interactions per 
event and of the number of reconstructed primary vertices.
Jets are required to have $\pT>\nobreak 20$~\GeV, 
and are reconstructed in the range  $|\eta| < 4.9$.   

Events are rejected if they include jets failing the quality criteria described in ref.~\cite{Aad:2011he}. 
To further reject spurious jet signals originating from cosmic rays or detector malfunctions, additional criteria are 
applied to the charged \pt fraction ($\chf$), 
defined as  the sum of the \pt of all tracks associated with the jet divided by the jet \pt, and to
the fraction of the jet energy contained in the electromagnetic layers of the calorimeter ($\emf$). 
Events are rejected if any of the two leading jets with $\pt >$ 100~\gev\  and $\abseta < 2.0$ satisfies either $\chf < 0.02$ or both 
$\chf < 0.05$ and $\emf>0.9$. To remove jets from additional $pp$ collisions, all jets with $\pt < 50$~GeV and $|\eta|<2.5$ are required to have at least one track identified as coming from the primary 
vertex. The primary vertex itself is defined as the vertex with the highest
summed track $\pt^2$.

Jets within the nominal acceptance 
of the ID ($|\eta| < 2.5$) and with $\pt > 20$~\GeV, are selected as originating from a \bquark (\btagged) 
if they satisfy requirements on the impact parameter
of the ID tracks, the secondary vertex reconstruction and the topology of $b$-
and $c$-hadron decays inside the jet. The \btagging algorithm~\cite{ATLAS:BtagCalib} uses a multivariate technique and, for this analysis, is configured to achieve an efficiency of 60\% 
for tagging \bjets in a MC sample of \ttbar events with corresponding rejection factors of 580, 8 and 23
 against jets originating from light quarks, $c$-quarks and $\tau$-leptons, respectively.

Electrons are reconstructed from cell-energy clusters in the
electromagnetic calorimeter matched to a track in the ID. Electron
candidates are required to have $\pt>7$~\GeV\ and $|\eta|<2.47$ and
must satisfy the  ``medium'' selection criteria described in
ref.~\cite{Aad:2011mk} and reoptimized for 2012 data. Electrons used to define the control regions are selected 
using the ``tight'' criteria, $\mathmbox{\pT>20\GeV}$, and with  an additional isolation requirement that the total transverse momentum of charged tracks within a cone of $\Delta R =$ 0.2 around 
the candidate be less than 10\% of the
reconstructed $\pT$.
Muon candidates are identified using a match between an extrapolated
ID track and one or more track segments in the muon
spectrometer~\cite{Aad:2010yt}, and are required to have $\pt>6$~\GeV\ and $|\eta| < 2.4$.  
Muons used to define the control regions are also required to have
$\pT > 20 \GeV$ and less than 1.8 $\GeV$ deposited in the calorimeter within a cone of $\Delta R =$ 0.2 
around the candidate. 

Following their reconstruction, candidate jets and leptons
may point to the same energy deposits in the
calorimeter. These overlaps are resolved by first discarding
any jet candidate within $\mathrm{\Delta} R=0.2$ of an electron candidate. Then, any electron or muon candidate within $\mathrm{\Delta} R = 0.4$ of any surviving jet is discarded. 

The missing transverse momentum, \ptmissvec, with magnitude \met, is constructed as the negative of the vector sum of the 
transverse momentum of all muons and electrons with $\pt > $10~GeV, jets with $\pt >$20~GeV, and
calibrated calorimeter energy clusters with $|\eta| < $4.9 not assigned to these objects~\cite{Aad:2012re}.

\section{Event selection}
\label{sec:fursel}
Two sets of signal regions are defined to provide sensitivity to the kinematic topologies associated with
different mass splittings  between the sbottom or the stop and the neutralino. 
In all cases, the presence of at least one primary vertex (with at least five associated 
tracks with $\pt > 0.4$ GeV) is required. 
Events are selected with $\met > 150$~\GeV\ and no electrons or muons identified in  the  final state.
For the signal region selections, jets within $|\eta|<2.8$ are ordered according to their \pt, with $n$ being their total number, and two jets are required to be \btagged. 
The following event-level variables are defined:
\begin{itemize}
\item[--] $\dphimin$ is defined as the minimum azimuthal distance, $\Delta \phi$, between any of the three leading jets and the $\ptmissvec$ vector
\begin{equation*}
\dphimin = \textrm{min}(|\phi_1 - \phi_{\ptmissvec}|,|\phi_2 - \phi_{\ptmissvec}|,|\phi_3 - \phi_{\ptmissvec}|).
\label{eq:dphi_def}
\end{equation*}
\noindent  Background multi-jet events are typically characterised by small values of \dphimin;

\item[--] \meff is defined as the scalar sum of the $\pt$ of the $k$ leading jets and the $\met$
\begin{equation*}
\meff(k) = \sum_{i=1}^{k} (\pt^{\rm jet})_i + \met,
\end{equation*}
where the index refers to the $\pt$-ordered list of jets;

\item[--] \HT3 is defined as the scalar sum of the $\pt$ of the $n$ jets, without including the three leading jets
\begin{equation*}
\HT3 = \sum_{i = 4}^{n} (\pt^{\rm jet})_i, 
\end{equation*}
\noindent where the index refers to the $\pt$-ordered list of jets;

\item[--] \mbb is defined as the invariant mass of the two \btagged jets in the event;

\item[--] \mct is the contransverse mass \cite{Tovey:2008ui} and is a kinematic variable that can be used to measure the masses of pair-produced
 semi-invisibly decaying heavy particles. For two identical decays of heavy particles into two visible
particles (or particle aggregates) $v_{1}$ and $v_{2}$, and
two invisible particles, $\mct$ is defined as:
\begin{equation*}
      \mct^{2}(v_{1},v_{2}) = \left[ E_{\rm T}(v_{1}) + E_{\rm T}(v_{2}) \right]^2 -  \left[  \textit{{\textbf p}}_{\rm T}(v_{1})
- \textit{{\textbf p}}_{\rm T}(v_{2}) \right]^2,
   \label{eq:mct_endpoints}
 \end{equation*}

\noindent where $E_{\rm T} = \sqrt{\pt^2+m^2}$. 
In this analysis, $v_1$ and $v_2$ are the two \bjets from the squark decays and the invisible particles are the two \neut particles or chargino decay products, depending on the assumption considered. 
The contransverse mass is an invariant under equal and opposite boosts of the parent particles in the transverse plane. For parent particles produced with small transverse boosts, \mct\ is 
bounded from above by an analytical combination of particle masses. This bound is saturated when the two 
visible objects are co-linear. 
For \ttbar\ events this kinematic bound is at 135~GeV.
For production of sbottom pairs the bound is given by:
\begin{equation*}
m_{\rm CT}^{\rm max} = \frac{m^2(\tilde{b}) - m^2(\neut)}{m(\tilde{b})}.
\label{eq:mct_max}
\end{equation*}
A similar equation can be written for production of stop pairs in terms of $m_{\tone}$ and $m_{\chipm}$. 
A correction to \mct\ for the transverse boost due to ISR is also applied~\cite{Polesello:2009rn}. 

\end{itemize}

\begin{table}[hpt!]
\footnotesize
\begin{center}\renewcommand\arraystretch{1.6}
\begin{tabular}{|l|c|c|}
\hline
\multirow{2}{*}{Description} & \multicolumn{2}{c|}{Signal Regions} \\
\cline{2-3}
  & \SRMCT & \SRISR \\
\hline
\hline
 Event cleaning &\multicolumn{2}{c|}{Common to all SR}\\ \hline
Lepton veto &\multicolumn{2}{c|}{No $e$/$\mu$ after overlap removal with $\pt$ $>$ 7(6) GeV for $e$($\mu$)}\\ \hline
\met  &$> 150$ \gev & $>250$~\gev \\ \hline
Leading jet $\pt(j_1) $ & $>$ 130 \gev & $> 150$~\gev  \\ \hline
Second jet $\pt(j_2) $ & $>$ 50 \gev,  & $>$ \lowptthr\ \gev  \\ \hline
Third jet $\pt(j_3) $ & veto if $>$ 50 \gev & $>$ \lowptthr\ \gev \\ \hline
$\Delta \phi (\ptmissvec,j_1) $ & - & $> 2.5$ \\ \hline
 {\btagging} & leading 2 jets & 2nd- and 3rd-leading jets \\
 & ($ \pt >$ 50 \gev, $|\eta| <$ 2.5) & ($\pt >$ \lowptthr\ \gev, $|\eta| <$ 2.5)\\ \cline{2-3}
 & \multicolumn{2}{c|}{$n_{b\mathchar`-\mathrm{jets}} = 2$} \\ \hline
$\dphimin$ & $>$ 0.4 & $>$ 0.4 \\ \hline
$\met/\meff(k)$ & $\met/\meff(2)>$ 0.25 &  $\met/\meff(3)>$ 0.25 \\ \hline
$ \mct $ & $>$ 150, 200, 250, 300, 350 \gev\ & - \\ \hline
\HT3 & - & $< 50 \gev$ \\ \hline
$m_{bb}$ & $>200 \gev$ & - \\ \hline
\end{tabular}
\caption{Summary of the event selection in each signal region.}\label{table:SRs}
\end{center}
\vspace{-0.5cm}
\end{table}

The definition of the two signal 
regions is summarised in table~\ref{table:SRs}. 
Signal region A (\SRMCT) targets signal events with large mass splitting between the squark and the neutralino, identifying 
 two \btagged high-\pt  
leading jets as products of the 
two sbottom or stop decays. 
Events are rejected if any other central ($|\eta|<2.8$) jets are found with $\pt > 50$~GeV. Multijet background is 
suppressed by selecting events with 
large \dphimin\ and $\met/\meff$.  The requirement $\mbb>200$~GeV is added to reduce backgrounds from production of top-quark (including top-quark pairs and single top-quark production processes), and  $Z$-bosons in association with heavy-flavour jets.  As a final selection criterion, five different thresholds on \mct ranging from 150~GeV to 350~GeV are applied. 
For a signal point corresponding to $m_{\bone}=500$~GeV and $m_{\neut}=1$~GeV, 3\% of the simulated events are retained by the \SRMCT selection with $\mct>250$~GeV.
   
Signal region B (\SRISR) is defined to enhance the sensitivity for a small squark--neutralino mass difference 
by explicitly selecting events with a high-\pt~jet, which is likely to have been produced as initial state radiation, 
recoiling against the squark-pair system. High thresholds on the leading jet \pt and on the missing transverse momentum, which are required to be almost back-to-back 
in $\phi$,  are imposed. The leading jet is required to be not \btagged, and two additional jets are required to be \btagged. As for \SRMCT, the multi-jet background is suppressed with appropriate selections on 
\dphimin\ and $\met/\meff$. A final upper requirement on the additional hadronic activity in the event, \HT3$<$50~GeV completes the selection for \SRISR. 
For a signal point corresponding to $m_{\bone}=300$~GeV and $m_{\neut}=270$~GeV, 10\% of the simulated events are retained by the \SRISR selection.

\section{Background estimate}
\newcommand{\captiondatamc}{The shaded band includes both the detector and theoretical systematic uncertainties. The SM prediction is normalised according to the MC expectations.}
\newcommand{\captionredarrow}{The red arrows indicate where a selection on the corresponding variable is applied.} 
The dominant SM background processes in the signal regions are the production of $W$ or  $Z$ bosons in association with
heavy-flavour jets (referred to as \WHF and \ZHF) and the production of top-quarks. 
Events with \ZHF production followed by $Z\rightarrow \nu\bar{\nu}$ decay have the same signature as the signal  and are the dominant background in \SRMCT. Top-quark (dominant in \SRISR) and \WHF 
production satisfy the signal region selections when a charged lepton is produced 
but the event is not rejected, either because the lepton is a hadronically decaying $\tau$, or because the electron or muon is not reconstructed. 
The dominant backgrounds are normalised in dedicated control regions (CRs) and then extrapolated to the signal regions using MC simulation. The control regions, detailed below, are 
defined by explicitly requiring the presence of one or two leptons (electrons or muons) 
in the final state together with further selection criteria similar to those of the corresponding signal regions. In particular, events with additional lepton candidates are vetoed applying the same lepton requirements used to veto events in the signal regions.
The single top-quark contribution accounts for 5\% to 20\% of the total top-quark background contribution, depending on the signal region considered, and is added to the $\ttbar$ background contribution with a relative 
normalisation corresponding to that predicted by the MC simulation, as described in section~\ref{sec:mc}.

The contributions from diboson and $\ttbar + W/Z$ processes are sub-dominant and they are collectively called ``Others'' in the following. They are estimated from MC simulation for both the 
signal and the control regions.

Finally, the background from multi-jet production is estimated from data using a procedure described in detail in 
ref.~\cite{Aad:2012fqa} modified to account for the flavour of the jets. 
The procedure consists 
of smearing the jet response in low-\etmiss\ seed events. The Gaussian core of the jet response function is 
obtained from dijet events, while the non-Gaussian tails are obtained from three-jet events, 
where the \met can be unambiguously attributed to the mis-measurement of one of the jets. The contribution from multi-jet production 
in the control regions is found to be negligible.

For \SRMCT, the contributions from top-quark, 
\Zjets and \Wjets production are estimated simultaneously with a profile likelihood fit to three control regions. 
For \SRISR it is difficult to identify a control region that probes the \Wjets background normalisation. Therefore, this contribution is estimated purely from MC simulation as described in section~\ref{sec:mc}, and only control regions for top-quark and \Zjets are defined.

A set of same-flavour opposite-sign two-lepton control regions with dilepton invariant mass near the $Z$ mass ($75< \mll  <105$~GeV) 
provides a data sample dominated by $Z$ production. 
For these control regions, labelled in the following as \CRASF and \CRBSF, the \pt of the leptons is added vectorially to the \ptmissvec to mimic the expected 
missing transverse momentum spectrum of $Z\rightarrow \nu \bar{\nu}$ events, and is indicated in the following as
$\met$ (lepton corrected). In addition, the \pt of the leading lepton is required to be above 90~GeV in order to further enhance 
the $Z$ production contribution. In the case of \CRASF, a $\mbb>200$~GeV selection is also imposed.

The set of control regions with exactly one lepton ($e,\mu$) 
in the final state provides a data sample dominated by top-quark and \Wjets production. 
A selection criterion is applied to the tranverse mass, $40\ \textrm{GeV} < \mt < 100\ \textrm{GeV}$, where \mt is 
defined as:
\begin{equation*}
\trmass = \sqrt{2 {p}_\mathrm{T}^{\rm lep} \etmiss - 2 {\textbf p}_\mathrm{T}^{\rm lep} \cdot \ptmissvec}
\end{equation*}
In the following, 
these control regions are labelled as \CRAL and \CRBL. \CRAL is used to estimate the contribution of the \Wjets background,  which is enhanced by the selection criterion $\mct>150$~GeV in \SRMCT. \CRBL is used to 
estimate the top-quark background in \SRISR.

To estimate top-quark production in \SRMCT, a different-flavour opposite-sign two-lepton control region (\CRADF) is defined requiring one electron and one muon in the final state with $m_{e \mu} >50$~GeV and \mct$>$75~GeV.

The definitions of the control regions are summarised in tables~\ref{tab:CRSRA} and~\ref{tab:CRSRB}.  

\begin{table}[ph!]
\footnotesize
\begin{center}\renewcommand\arraystretch{1.6}
\begin{tabular}{|c|c|c|}
\hline
\CRAL & \CRASF & \CRADF \\ \hline
\hline
One $e$ or $\mu$ & $e^\pm e^\mp$ or $\mu^\pm\mu^\mp$ & $e^\pm\mu^\mp$ \\ \hline
\multicolumn{3}{|c|}{Veto additional lepton candidates ($\pt(e)>$7 GeV $\pt(\mu)>$6 GeV)} \\ \hline 
\multicolumn{3}{|c|}{Only two reconstructed jets with $\pt > 50$~GeV} \\ \hline 
$\pt(j_1)  > 130\ \mathrm{GeV}$  & $\pt(j_1) > 50\ \mathrm{GeV}$ & $\pt(j_1) > 130\ \mathrm{GeV}$ \\
$\pt(j_2) > 50\ \mathrm{GeV}$ & $\pt(j_2) > 50\ \mathrm{GeV}$ & $\pt(j_2) > 50\ \mathrm{GeV}$ \\
$\etmiss > 100$ GeV & $\etmiss(\rm lepton\mathchar`-corrected) > 100$ GeV & $\etmiss > 100$ GeV \\ \hline
\multicolumn{3}{|c|}{Two reconstructed $b$-jets ($\pt > 50$)}\\ \hline
$40\ \mathrm{GeV} < \mt < 100$ GeV & $75\ \mathrm{GeV}< \mll < 105$ GeV & $\mll > 50$ GeV \\ \hline
$\mct>150$~GeV &  lepton $\pt > 90\ \mathrm{GeV}$ & $\mct>75$~GeV \\ \hline
 --  & $\mbb>200$~GeV & -- \\
\hline
\end{tabular}
\caption{\label{tab:CRSRA} Definitions of the three \SRMCT control regions.}
\end{center}
\end{table}
 
\begin{table}
\footnotesize
\begin{center}\renewcommand\arraystretch{1.6}
\begin{tabular}{|c|c|}
\hline
\CRBL & \CRBSF \\ \hline
\hline
One $e$ or $\mu$ & $e^\pm e^\mp$ or $\mu^\pm \mu^\mp$ \\ \hline
\multicolumn{2}{|c|}{Veto additional lepton candidates ($\pt(e)>$7 GeV $\pt(\mu)>$6 GeV)} \\ \hline
\multicolumn{2}{|c|}{Only three reconstructed jets with $\pt > \lowptthr$~GeV} \\ \hline 
$\pt(j_1) > 130\ \mathrm{GeV}$ & $\pt(j_1) > 50\ \mathrm{GeV}$  \\
$\etmiss > 120$ GeV & $\etmiss(\rm lepton\mathchar`-corrected) > 100$ GeV \\ \hline
\multicolumn{2}{|c|}{$j_1$ anti $b$-tagged; $j_2$ and $j_3$ $b$-tagged}\\ \hline
$40\ \mathrm{GeV} < \mt < 100$ GeV & $75\ \mathrm{GeV}< \mll < 105$ GeV \\ \hline
-- & Lepton $\pt > 90\ \mathrm{GeV}$ \\ \hline
\multicolumn{2}{|c|}{$\HT3 <50$~GeV}\\ 
\hline
\end{tabular}
\caption{\label{tab:CRSRB} Definitions of the two \SRISR control regions.}
\end{center}
\end{table}

The distribution of the \mct variable in \CRADF (before the \mct selection) and of the leading lepton \pt (before the \pt selection) in \CRASF are shown before the fit in figure~\ref{fig:CRSRA}. Similarly, the transverse mass distribution of the leading-lepton--\met system in \CRBL (before the \mt selection) and the invariant mass distribution of the two leptons in \CRBSF (before the \mll selection) are shown in figure~\ref{fig:CRSRB}. 
In these figures the data set used corresponds to an integrated luminosity of 20.3~\ifb\ and  the normalisations described in section~\ref{sec:mc} are assumed.

\begin{figure}[!htb]
\includegraphics[width=0.45\columnwidth]{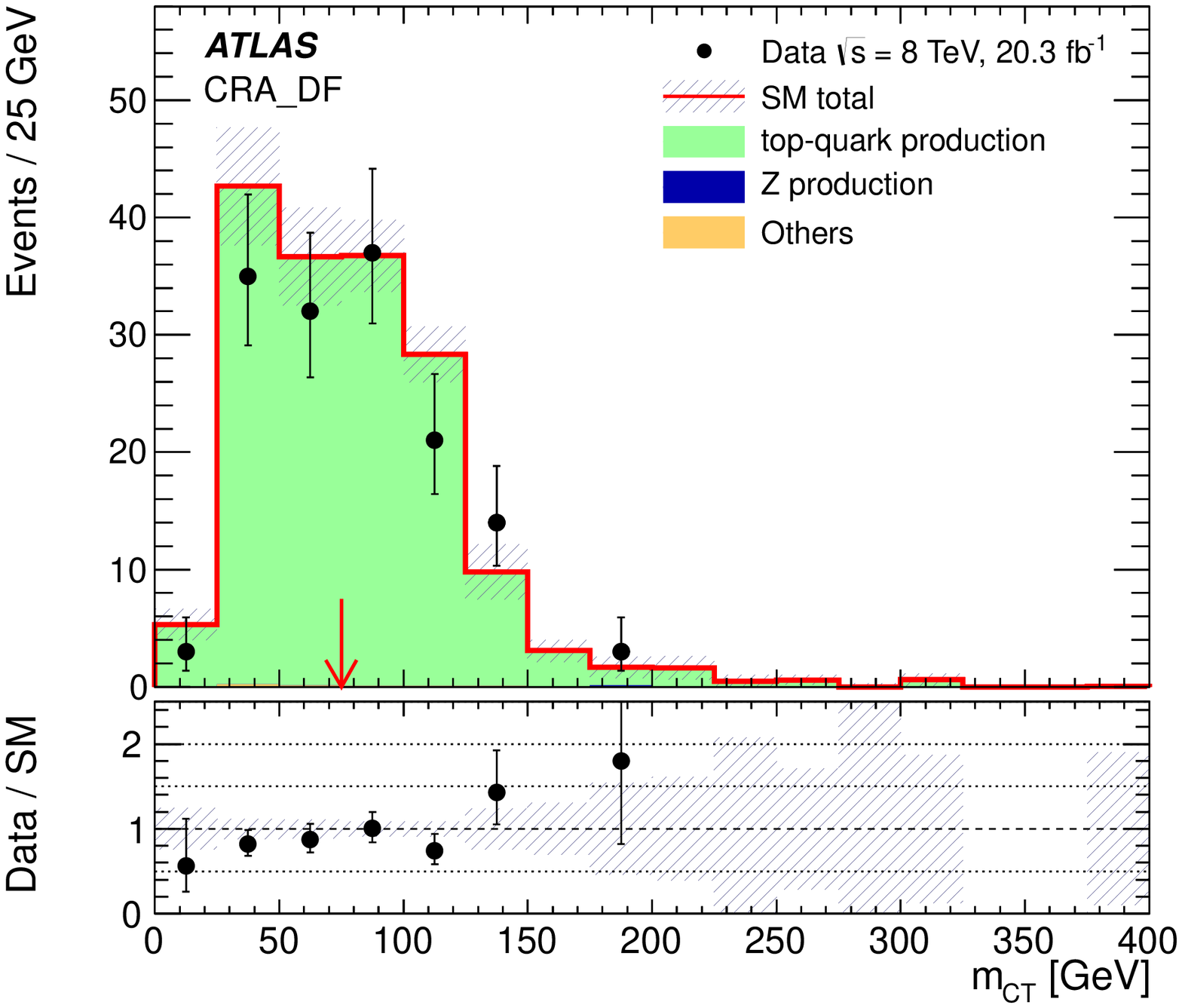}
\includegraphics[width=0.45\columnwidth]{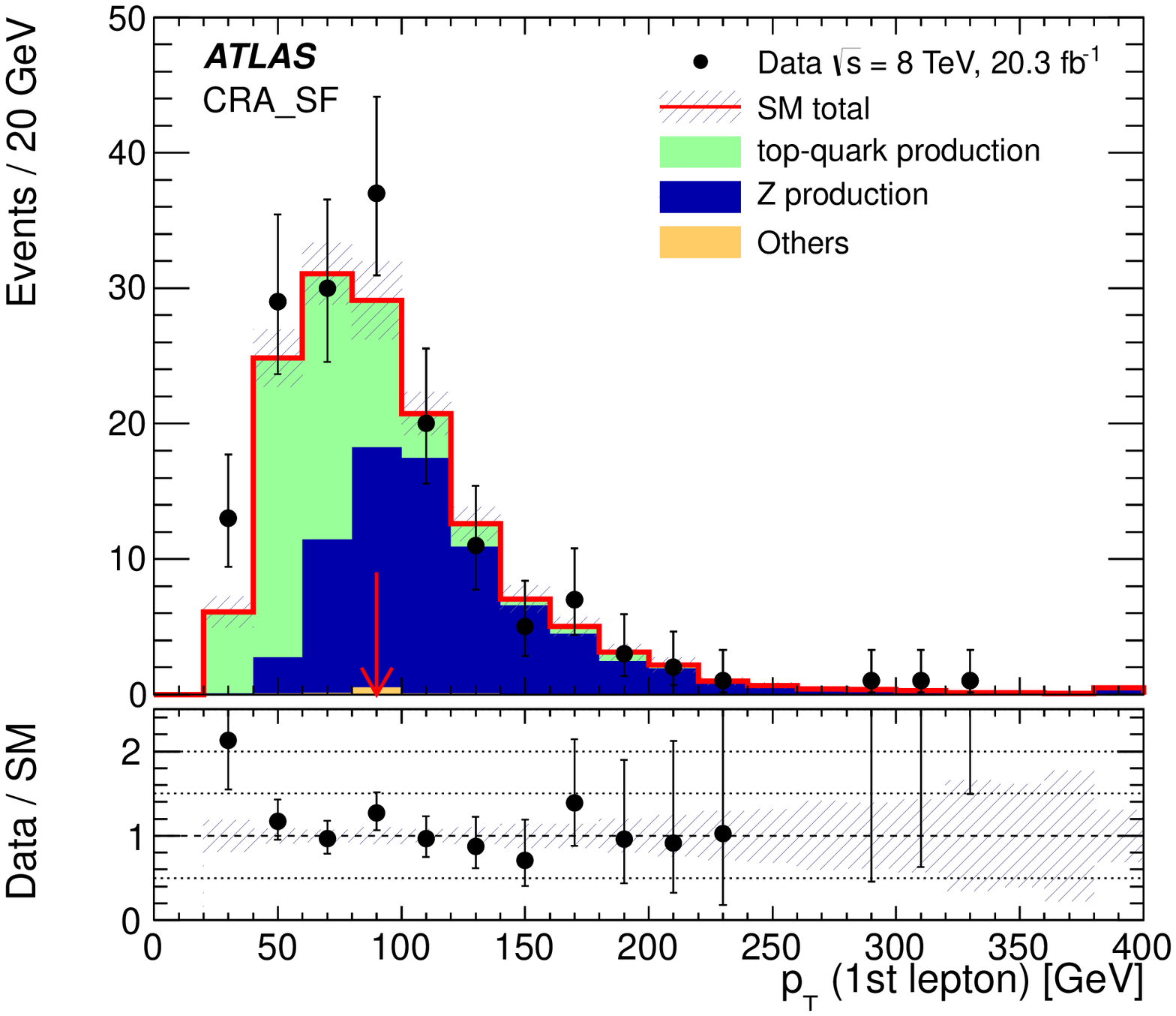}
\caption{Left: \mct distribution in \CRADF omitting the requirement on the \mct variable. Right: leading lepton \pt distribution in \CRASF with all the selections applied except the requirement on this variable. \captionredarrow\ \captiondatamc\ The last bin in each histogram contains the integral of all events with values greater than the upper axis bound.}
\label{fig:CRSRA}
\end{figure}

\begin{figure}[!htb]
\includegraphics[width=0.45\columnwidth]{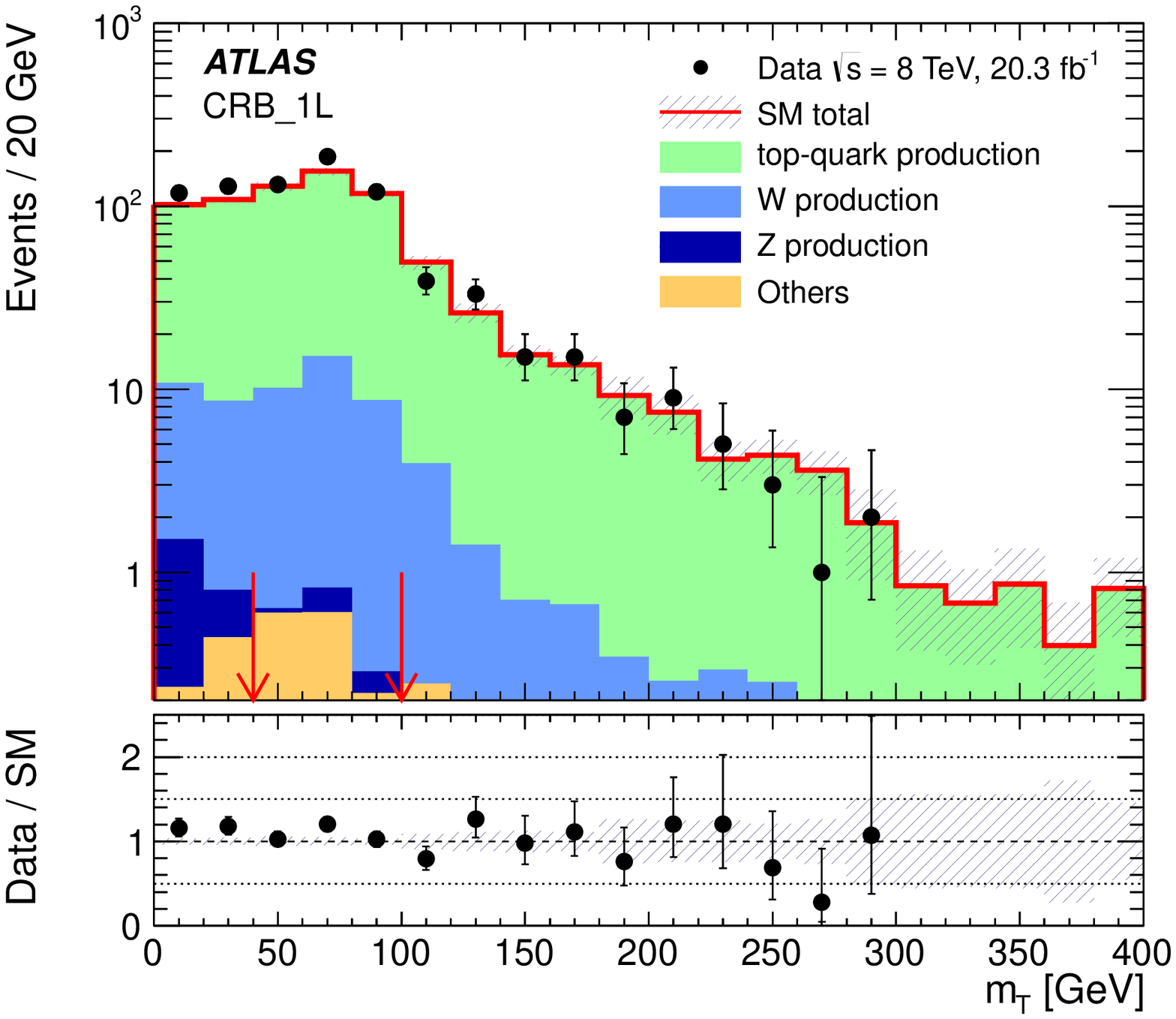}
\includegraphics[width=0.45\columnwidth]{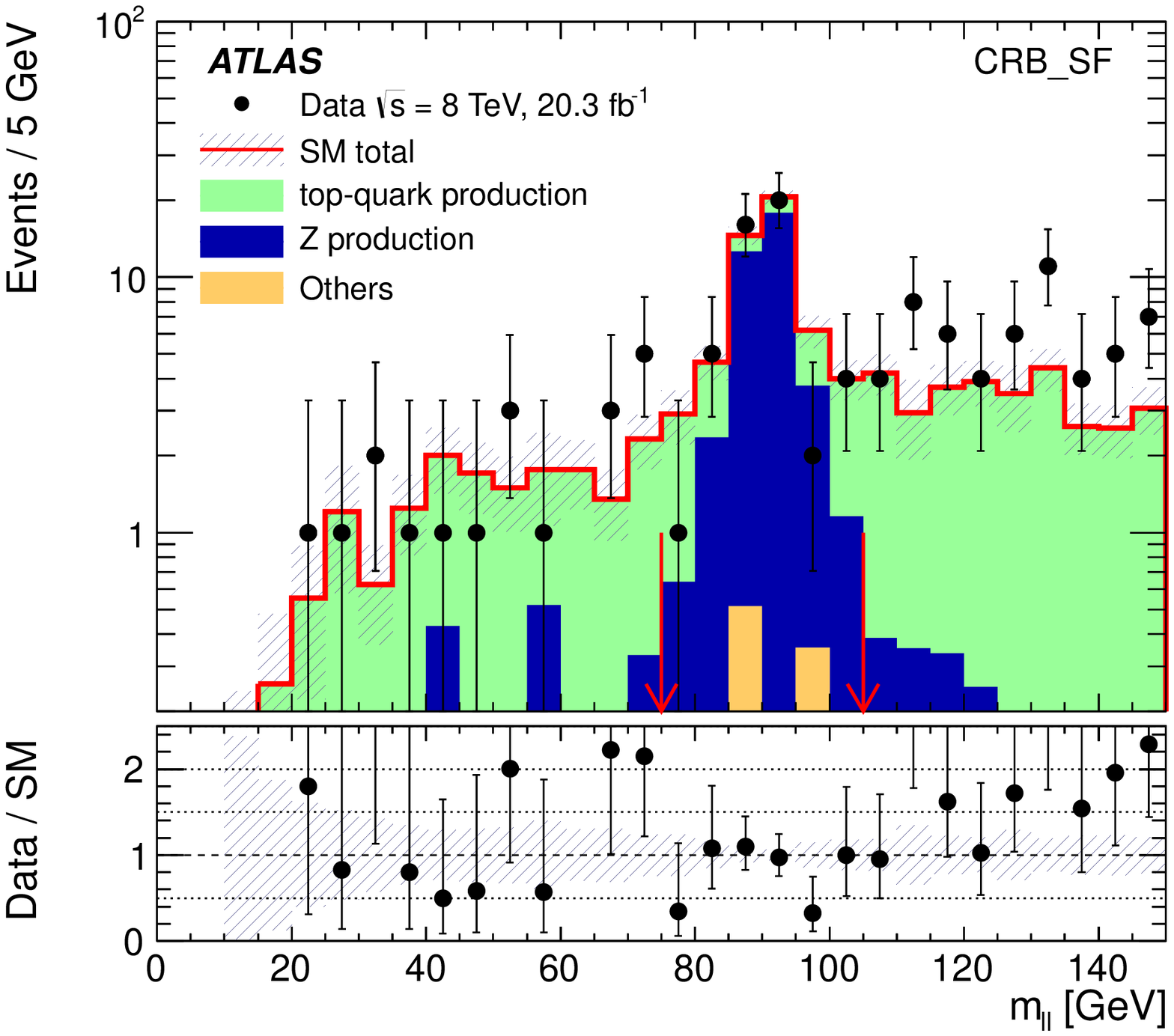}
\caption{Left: transverse mass distribution of the leading-lepton--\etmiss system in \CRBL omitting the \mt requirement. Right: dilepton invariant mass distribution in \CRBSF omitting the \mll requirement. \captionredarrow\ \captiondatamc}
\label{fig:CRSRB} 
\end{figure}

The observed numbers of events in the various CRs are used to generate internally consistent SM background estimates for each of the SRs via a profile likelihood fit. This procedure takes 
into account CR 
correlations due to common systematic uncertainties as well as contaminations from other SM processes and/or SUSY signal events, when a particular model is considered for exclusion. Systematic uncertainties, discussed in detail in section~\ref{sec:systematics}, 
are treated as nuisance parameters in the fit and are constrained with Gaussian functions 
taking into account correlations between sample estimates. 
The likelihood function is built as the product of Poisson probability functions, describing the observed and expected number 
of events in the control and (when excluding SUSY models) signal regions, and the constraints on the nuisance parameters. As a result, the impact of some of the systematic uncertainties that are correlated between the CRs and the corresponding SR is reduced.

The free parameters of the fit are the overall normalisation values of the top-quark, $W$+jets and $Z$+jets processes for \SRMCT, and of the top-quark and 
$Z$+jets processes for \SRISR. The contributions from all other background processes are fixed at the values expected from MC. The fit results in the control regions are summarised in tables~\ref{tab:fitresSRA} and~\ref{tab:fitresSRB} for \SRMCT and \SRISR, respectively. These results are found to be compatible with MC yields predicted before the fit, which are also given in the tables. 

The predictions from the fit are in good agreement with the MC estimates with a maximum discrepancy of two 
standard deviations observed in one of the control regions.

The reliability of the MC extrapolation of the SM background estimate outside of the control regions is evaluated in 
several validation regions. The first set of validation regions is defined with the same 
kinematic selection as the control regions but requiring only one jet to be \btagged. 
They are used to verify the performance of the $b$-tagging algorithm in a larger sample of events. 
A second set of validation regions is defined with the same selection criteria as the signal regions, but with one of  
the requirements inverted. In all cases these validation regions are background-dominated with a potential signal 
contamination of less than 20\% for the signal models considered. 
For \SRMCT, two validation regions are explored by imposing either $\mct < 100$ GeV or $\mbb < 200$ GeV. 
To validate \SRISR, a validation region with the selection $\HT3 > 50$~GeV is defined as well as a second validation region 
with two leptons of different flavour to verify the normalisation of the top-quark background.

Good agreement is found in each case, with a difference of less than one standard deviation between the expectations and the number of 
observed events.

\begin{table}
\begin{center}
\begin{tabular}{lccc}
\hline
{\bf Channel}  & {\bf \CRAL} & {\bf \CRASF} & {\bf \CRADF} \\
\hline
Observed events          & $135$              & $68$              & $75$                    \\
\hline
\multicolumn{4}{c}{Fitted background events}\\
\hline
 Total SM        & $135 \pm 11$          & $68 \pm 8$          & $75 \pm 9$ \\ 
\hline
Top-quark production &   $91 \pm 17$          & $10.0 \pm 1.3$          & $75 \pm 9$      \\
$Z$ production &   $0.46 \pm 0.12$          & $58 \pm 8$          & $0.07_{-0.07}^{+0.10}$              \\
$W$ production     & $40 \pm 20$          & $<0.1$          & $0.06 \pm 0.03$              \\
Others            &  $3.8 \pm 2.0$          & $0.44 \pm 0.18$          & $0.37 \pm 0.13$              \\
\hline
\multicolumn{4}{c}{MC expected events}\\
\hline
\noalign{\smallskip}\noalign{\smallskip}
Top-quark production         & $100$          & $11.0$          & $82$              \\
$Z$ production         & $0.44$          & $54$          & $0.07$              \\
$W$ production         & $42$          & $<0.1$          & $0.07$              \\
Others         & $3.8$          & $0.44$          & $0.37$              \\
\hline

\end{tabular}
\caption{\label{tab:fitresSRA} Results of the fit for the control regions adopted for \SRMCT. Expected yields derived from MC simulation using theoretical cross sections are also shown. Uncertainties quoted include statistical and detector-related systematic effects. The central values of the fitted sum of backgrounds in the control regions agree with the observations by construction. The uncertainty on the total background estimate can be smaller than some of the individual uncertainties due to anticorrelations.}
\end{center}
\end{table}

\begin{table}
\begin{center}
\begin{tabular}{lcc}
\hline
{\bf Channel}  & {\bf \CRBL} & {\bf \CRBSF}\\ \hline
Observed events          & $437$              & $48$                    \\\hline
\multicolumn{3}{c}{Fitted background events}\\
\hline
Total SM         & $437 \pm 21$          & $48 \pm 7$              \\\hline
        Top-quark production         & $403 \pm 27$          & $16.2 \pm 2.2$              \\
        $Z$ production         & $0.26 \pm 0.15$          & $31 \pm 7$              \\
        $W$ production         & $32 \pm 20$          & $<0.1$              \\
        Others         & $1.4 \pm 0.5$          & $1.0 \pm 0.5$              \\
\hline
\multicolumn{3}{c}{MC expected events}\\
\hline
\noalign{\smallskip}\noalign{\smallskip}
        Top-quark production         & $370$          & $15$              \\
        $Z$ production         & $0.32$          & $38$              \\
        $W$ production         & $32$          & $<0.1$              \\
        Others                 & $1.4$          & $1.0$              \\

\hline
\end{tabular}
\caption{\label{tab:fitresSRB} Results of the fit for the control regions adopted for \SRISR. Expected yields as derived from MC using theoretical cross sections are also shown. Uncertainties quoted include statistical and detector-related systematic effects. The central values of the fitted sum of backgrounds in the control regions agree with the observations by construction. The uncertainty on the total background estimate can be smaller than some of the individual uncertainties due to anticorrelations. The $W$ production estimate is normalised using the nominal theoretical cross section and with the associated uncertainties discussed in section~\ref{sec:systematics}.}
\end{center}
\vspace{-0.8cm}
\end{table}

\section{Systematic uncertainties}
\label{sec:systematics}
The dominant detector-related systematic effects are due to the uncertainties on the 
jet energy scale (JES) and resolution (JER), and on the $b$-tagging efficiency. The impact of these uncertainties is reduced through the normalisation of the dominant backgrounds in the control regions with kinematic selections resembling those of the corresponding signal region. 

The JES uncertainty is determined using the techniques described in refs.~\cite{Aad:2011he,Aad:2012vm}, leading to a slight dependence on \pt, $\eta$, jet flavour, number of 
primary vertices and proximity to adjacent jets. The JER uncertainty is obtained from in-situ measurements of the jet response asymmetry in dijet events~\cite{Aad:2011he}. These 
uncertainties on jets are propagated to the \met\ measurement, and 
additional uncertainties on \met\ arising from energy deposits not associated with any reconstructed objects are also included. The relative impact on the event yields from the JES (JER) uncertainty is between 1--5\% (1--8\%) in the different \SRMCT regions and is 3\% (8\%) in \SRISR. 

The $b$-tagging uncertainty is evaluated by varying the $\pt$- and flavour-dependent scale factors applied to each jet in the simulation within a 
range that reflects the systematic uncertainty on the measured tagging efficiency and rejection rates. The relative impact of this uncertainty on the final event yield is dominated by the uncertainty in the \btagging efficiency. The uncertainty amounts to 2--10\% in the different \SRMCT regions and 2\% in \SRISR.

In the case of \SRISR, an uncertainty is also associated with the requirement that  jets with $\pt<50$~GeV have at least one track originating from the primary vertex. It has a relative impact on the final event yields of 7\%. Other detector systematic uncertainties like pile-up or trigger effects are found to have negligible impact on the analysis.

Theoretical uncertainties on the modeling of the $\ttbar$+jets background are assessed. The uncertainty due to 
the choice of the MC generator is estimated by comparing the predictions of the \powheg and \alpgen generators, both interfaced to the \herwigjimmy parton shower (PS) and hadronisation (HAD) calculations. 
The PS/HAD uncertainty is estimated by comparing samples generated with \powheg interfaced 
to either \pythia or \herwigjimmy. The uncertainty due to the ambiguity in the  renormalisation and factorisation scales is 
estimated by individually doubling or halving them. 
The uncertainty in the ISR and final-state radiation (FSR) is estimated 
by comparing {\sc AcerMC} samples generated with different amounts of ISR/FSR, as constrained by recent ATLAS measurements~\cite{ATLAS:2012al}. 
The PDF uncertainties are derived by varying the 52 PDFs in the CT10 NLO error set following the Hessian method and rescaling to the 68\% confidence level. Since \ttbar and single top-quark productions are treated together, these theoretical uncertainties are also assigned to the single top contribution and it was checked that they account for the difference in event kinematics between the samples. The final relative uncertainties on the yields due to theoretical uncertainties in top-quark production modeling are less than 0.7\% in \SRMCT and 4\% in \SRISR, dominated mostly by PS/HAD and scale uncertainties.

Uncertainties in the $W/Z$+jets simulation are evaluated by comparing the predictions of the \sherpa and \alpgen generators and, by varying the \sherpa scales related to the matching scheme, the strong coupling constant, and the renormalisation and factorisation scales. The PDF uncertainties are evaluated following the same procedure as for the top-quark background. For \SRISR, $W$+jets production is estimated from MC simulation and an additional uncertainty of 26\% due to the \WHF contribution is included~\cite{Aad:2013vka}. 
The relative uncertainties on the yields are in the range 0.3--3.1\% in \SRMCT, mostly dominated by the difference between \sherpa and \alpgen. For \SRISR, a 3\%  uncertainty  is assigned to \ZHF and 9\% to \WHF, the latter uncertainty being dominated by the MC normalisation.

An uncertainty of 100\% is derived for the multi-jet prediction 
by studying variations of the resolution function. Finally,  
uncertainties of  
30\%  and  of 50\% are assigned to the cross sections of \ttW and of \ttZ 
production, respectively~\cite{Garzelli:2012bn,Campbell:2012dh}.

\section{Results and interpretation}
\newcommand{\captiondatamcfit}{The shaded band includes statistical, detector-related and theoretical systematic uncertainties. The backgrounds are normalised to the values determined in the fit.}
\newcommand{\overflowcaption}{The rightmost bin in the figures includes the overflows.}

The number of data events 
observed in each signal region is reported in table~\ref{tab:0lepFinalResultsSR_2jX}, together with the SM background expectation after the fit. 
Figures~\ref{fig:res_SRA} and~\ref{fig:res_SRB} show the comparison between the SM 
prediction and the observed data for some relevant kinematic distributions in \SRMCT and \SRISR, respectively. An example of a SUSY process with a large mass difference between the squark and the lightest neutralino is also shown for reference in each case.

\begin{table}
\footnotesize
\begin{center}
\begin{tabular}{lcccccc}
\hline
{\bf Channel}  & \multicolumn{5}{c}{{\bf \SRMCT, \mct selection}}  & {\bf \SRISR} \\
\hline
  & $150$ GeV  & $200$ GeV & $250$ GeV &  $300$ GeV  &  $350$ GeV &   \\
\hline
Observed & 102 & 48 & 14 & 7 & 3 & 65 \\
\hline
Total SM & $94\pm13$ & $39\pm 6$ & $15.8\pm2.8$ & $5.9\pm1.1$ & $2.5\pm0.6$ & $64 \pm 10$ \\
\hline
Top-quark  & $11.1 \pm 1.8$          & $2.4 \pm 1.4$          & $0.44 \pm 0.25$          & $<0.01$          & $<0.01$ & $41\pm 7$  \\
$Z$ production  &  $66 \pm 11$          & $28 \pm 5$          & $11.4 \pm 2.2$          & $4.7 \pm 0.9$          & $1.9 \pm 0.4$  & $13\pm 4$  \\
$W$ production & $13 \pm 6$          & $4.9 \pm 2.6$          & $2.1 \pm 1.1$          & $1.0 \pm 0.5$          & $0.46 \pm 0.26$ & $8\pm 5$  \\
Others  & $4.3 \pm 1.5$          & $3.4 \pm 1.3$          & $1.8 \pm 0.6$          & $0.12 \pm 0.11$          & $0.10_{-0.10}^{+0.12}$ & $2.0\pm 1.0$  \\
Multijet & $0.21 \pm 0.21$  & $0.06 \pm 0.06$          & $0.02 \pm 0.02$                  & $<0.01$                  & $<0.01$ & $0.16\pm 0.16$  \\

\hline
\end{tabular}
\caption{For each signal region, the observed event yield is compared with the background prediction obtained from the fit. Statistical, detector-related and theoretical systematic uncertainties are included, taking into account correlations. 
\label{tab:0lepFinalResultsSR_2jX}}
\end{center}
\vspace{-0.5cm}
\end{table}

\begin{figure}[!htb]
\begin{center}
\includegraphics[width=0.49\columnwidth]{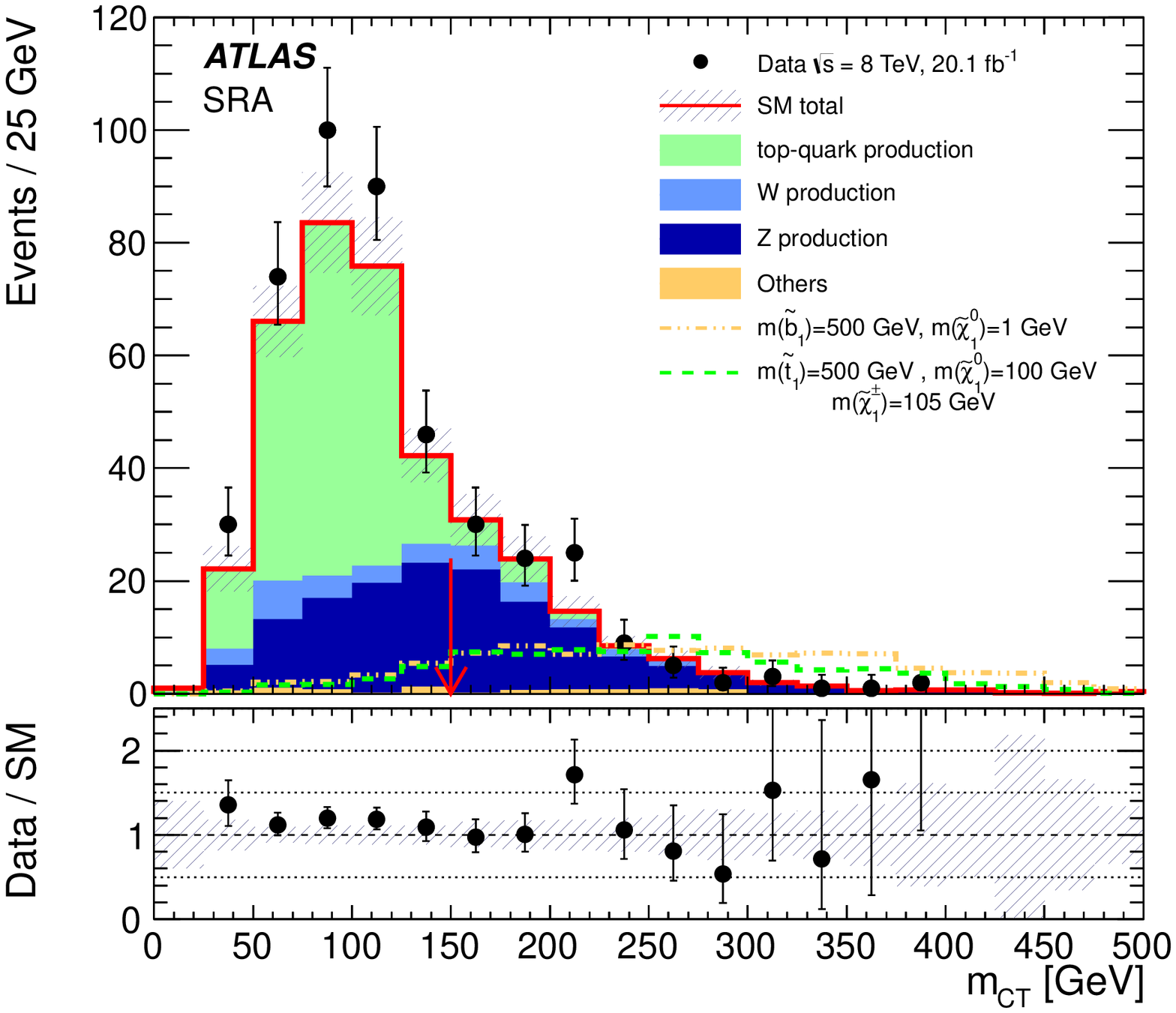}
\includegraphics[width=0.49\columnwidth]{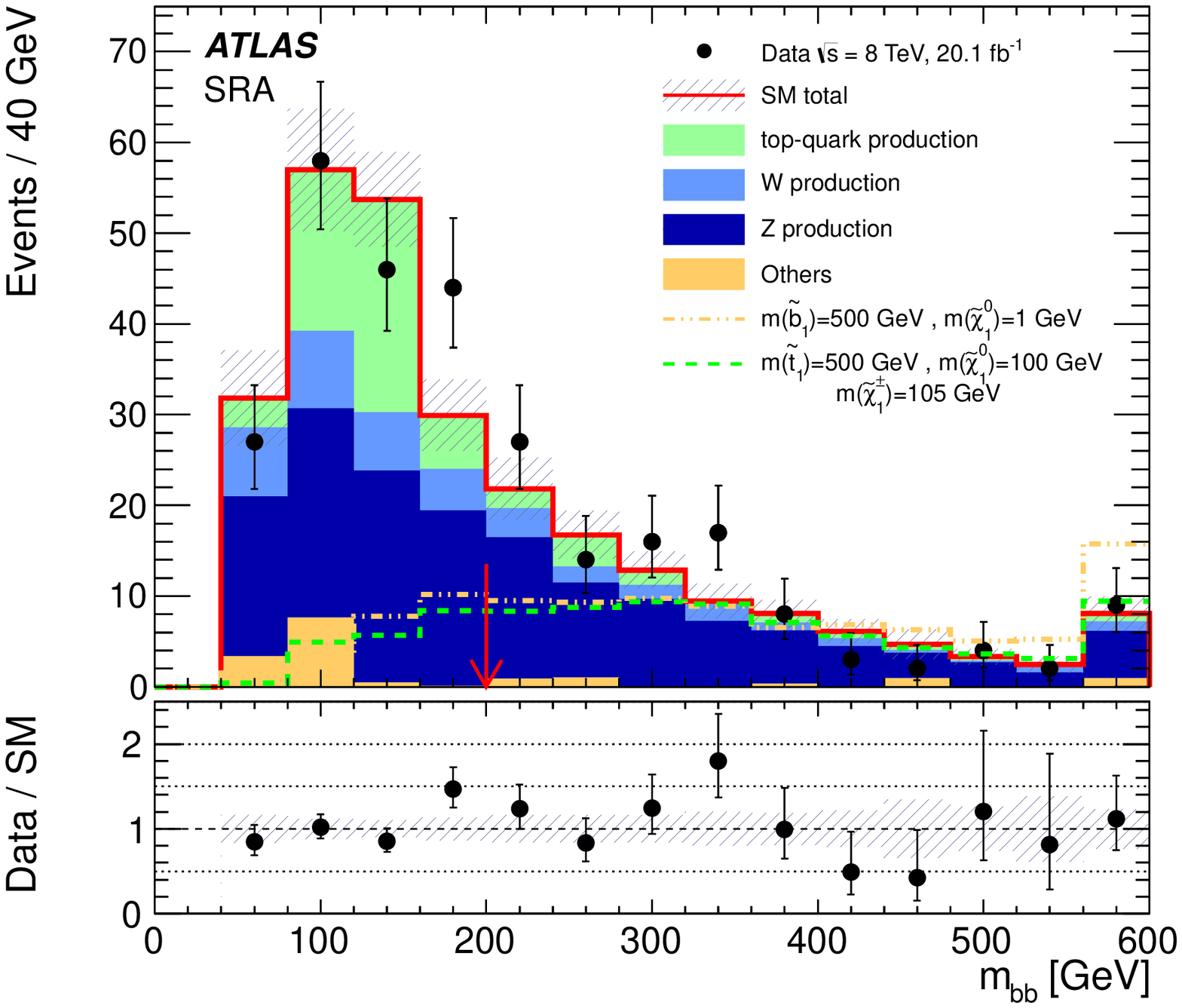}
\end{center}
\vspace{-0.3cm}
\caption{Left: \mct~distribution in \SRMCT with all the selection criteria applied except the \mct thresholds. Right: \mbb distribution in \SRMCT with all selection criteria applied including $\mct>150$~GeV. \captiondatamcfit\ \captionredarrow\ 
For illustration the distributions expected for two signal models are displayed. 
The models correspond to $m_{\bone}=500$~GeV and $m_{\neut}=1$~GeV (orange dash-dot line) 
and $m_{\tone}=500$~GeV and $m_{\neut}=100$~GeV (green dash line). \overflowcaption} 
\label{fig:res_SRA}
\vspace{-0.2cm}
\end{figure}

\begin{figure}[!htb]
\begin{center}
\includegraphics[width=0.49\columnwidth]{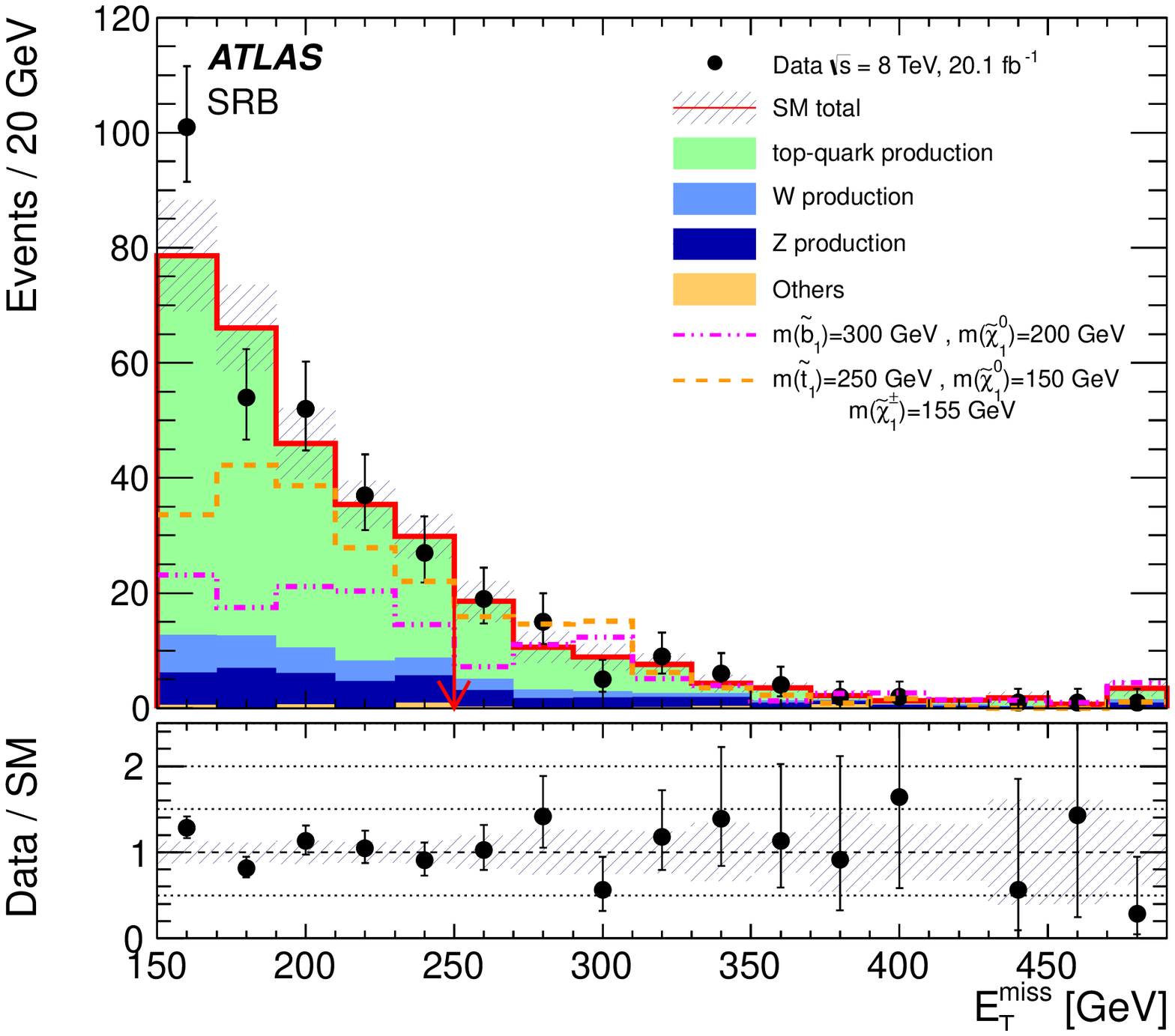}
\includegraphics[width=0.49\columnwidth]{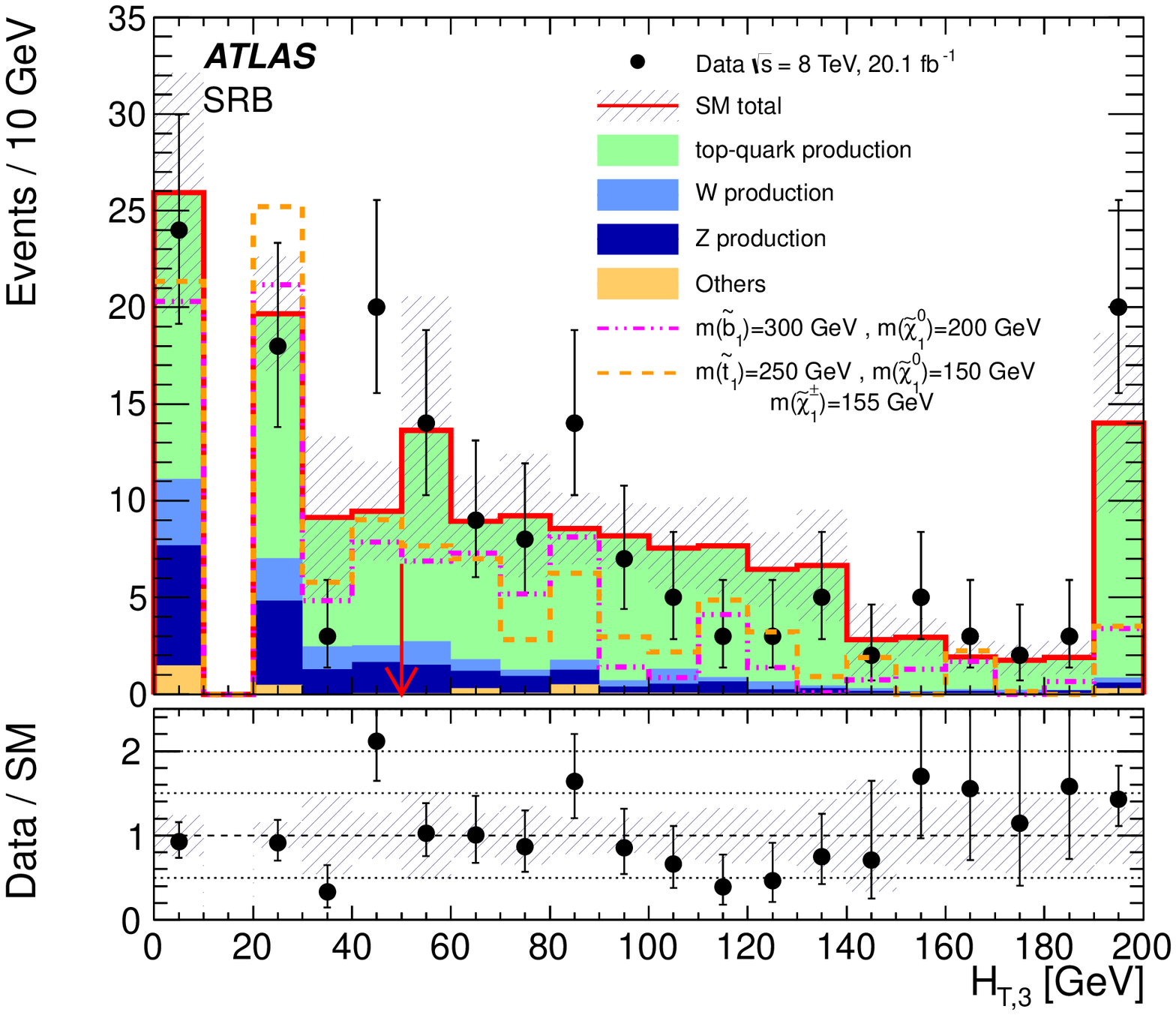}
\end{center}
\vspace{-0.3cm}
\caption{Left: \met~distribution with all \SRISR selection criteria applied except the final \met requirement. 
Right: distribution of \HT3 with all \SRISR selection criteria applied except the \HT3 requirement. Since jets have $\pt>20$~GeV by 
construction, the leftmost bin 
contains events where no additional jets are 
present (\HT3=0~GeV) while the second bin is empty. 
\captiondatamcfit\ \captionredarrow\ 
For illustration the distributions expected for two signal models are displayed.
The models correspond to $m_{\bone}=300$~GeV and $m_{\neut}=200$~GeV (purple dash-dot line) 
and $m_{\tone}=250$~GeV and $m_{\neut}=150$~GeV (orange dash line). \overflowcaption}
\label{fig:res_SRB}
\vspace{-0.2cm}
\end{figure}

No significant excess above the SM expectation is observed in any of the signal regions. The results are used to 
derive upper limits on the number of beyond the SM (BSM) events for each 
signal region, 
assuming no systematic uncertainties for these events and neglecting any possible contamination in the control regions. Scaling by the integrated luminosity, these can be interpreted as a corresponding upper limit on the cross section, $\sigma_{\mathrm{vis}}$, defined as 
\begin{equation*}
\sigma_{\mathrm{vis}} = \sigma \cdot A \cdot \epsilon
\end{equation*}
\noindent where $\sigma$, $A$ and $\epsilon$ are, respectively, the production cross section, the acceptance and the 
selection efficiency for a BSM signal. The 95\% confidence level (CL) limits are computed using the \CLs prescription~\cite{Read:2002hq}.
Table~\ref{tab:ULlimits} summarises, for each signal region,  the estimated SM background yield, the observed numbers of events, 
and the expected and observed upper limits on event yields from a BSM signal and on $\sigma_{\mathrm{vis}}$.

\begin{table}
\footnotesize
\begin{center}\renewcommand\arraystretch{1.6}
\begin{tabular}{|c|c|c|c|c|c|c|}
\hline

Signal Regions & Background & Observed & \multicolumn{4}{c|}{95\% CL upper limit on} \\\cline{4-7}
& estimate & data & \multicolumn{2}{c|}{BSM event yield} & \multicolumn{2}{c|}{$\sigma_{\mathrm{vis}}$ (fb)}  \\\cline{4-7}
                & & & exp. & obs. & exp. & obs.\\
\hline
\hline
\SRMCT ($\mct>150$ GeV) & $94 \pm 13$  & 102 & ${32}^{+13}_{-9}$ & $38$ & $1.6^{+0.6}_{-0.4}$ & $1.9$\\
\SRMCT ($\mct>200$ GeV) & $39 \pm 6$ & 48 & ${19}^{+8}_{-5}$ & $26$ & $0.94^{+0.40}_{-0.25}$ & $1.3$\\
\SRMCT ($\mct>250$ GeV) & $15.8\pm2.8$ & 14 & ${10.2}^{+4.6}_{-3.0}$ & $9.0$ & $0.51^{+0.22}_{-0.14}$ & $0.45$\\
\SRMCT ($\mct>300$ GeV) & $5.9 \pm 1.1$ & 7 & ${6.5}^{+3.3}_{-2.1}$ & $7.5$ & $0.32^{+0.16}_{-0.1}$ & $0.37$\\
\SRMCT ($\mct>350$ GeV) & $2.5 \pm 0.6$ & 3 & ${4.7}^{+2.6}_{-1.6}$ & $5.2$ & $0.23^{+0.13}_{-0.08}$ & $0.26$\\
\SRISR & $64 \pm 10$ & 65 & ${26}^{+10}_{-7}$ & $27$ & $1.21^{+0.45}_{-0.35}$ & $1.3$\\
\hline
\end{tabular}
\caption{Expected and observed event yields with the corresponding upper limits  on BSM signal yields and $\sigma_{\mathrm{vis}} = \sigma \cdot A \cdot \epsilon$ for all the signal regions defined. }\label{tab:ULlimits}
\end{center}
\vspace{-0.5cm}
\end{table}

The results are interpreted in various SUSY scenarios assuming a SUSY particle mass 
hierarchy such that the lightest third-generation squark decays exclusively via $\bone \rightarrow b\neut$ or $\tone \rightarrow b \chipm$ for production of sbottom or stop pairs, respectively. For the latter case, two different values of 
$\Delta m$ between the lightest chargino and neutralino are probed to assess the impact of the lepton and jet vetoes applied in the analysis.
 
Systematic uncertainties on the signal acceptance include experimental uncertainties, mostly dominated by \btagging (20--30\% in \SRMCT, $\sim 15$--30\% in \SRISR) and JES (4--30\% in \SRMCT, 20--40\% in \SRISR) uncertainties. These uncertainties are assumed to be fully correlated with those of the background.

For \SRISR, the uncertainties due to the modelling of ISR processes are assessed by changing the strength of the parton shower controlled by \pythia and by doubling and halving the values of the following three parameters: (i) the factorisation and renormalisation scales; (ii) the matching distance between a parton and a jet; (iii) the scale at which $\alpha_S$ is evaluated at every parton radiation step. The relative changes due to each of these individual variations are assumed to be uncorrelated and are added in quadrature. The overall uncertainty due to ISR  depends on the mass difference between the squark and the LSP, with a maximum value of 30\% on the signal acceptance when the mass difference is on the order of 10~GeV and quickly dropping down to a plateau of 7--10\% for mass differences above 25~GeV. 
This uncertainty has a negligible dependence on the squark mass for the
mass range considered in this analysis.

Figure~\ref{fig:finallimit} shows the observed (solid lines) and expected (dashed lines) exclusion limits for the sbottom pair production scenario obtained by taking, for each signal mass configuration, the signal region with the best expected limit. These limits are obtained using a likelihood test which compares the observed numbers of events in the signal regions with the fitted background expectation and the ensuing signal contamination in the corresponding CRs for a given model. Sensitivity to scenarios with large mass difference ($>100$~GeV) between the \bone and the \neut\ is achieved with the successive \mct thresholds used in \SRMCT. Sensitivity to scenarios with smaller mass differences is achieved predominantly with the dedicated \SRISR selection.
Sbottom masses up to 620~GeV are excluded at 95\% CL for $m_{\neut} < 120$~GeV. Differences in mass above 50~GeV between the \bone and the \neut are excluded up to sbottom masses of 300~GeV.
If the branching ratio of $\bone \rightarrow b\neut$ is reduced to 60\% and assuming that the analysis is not sensitive to other possible decays, the excluded upper limit on the 
sbottom mass for $m_{\neut} < 150$~GeV is reduced to 520~GeV. Similarly for $m_{\bone}=250$~GeV, the upper limit on $m_{\neut}$ is reduced by 30~GeV.

\begin{figure}[!hpt]
\begin{center}
   \includegraphics[width=0.65\columnwidth]{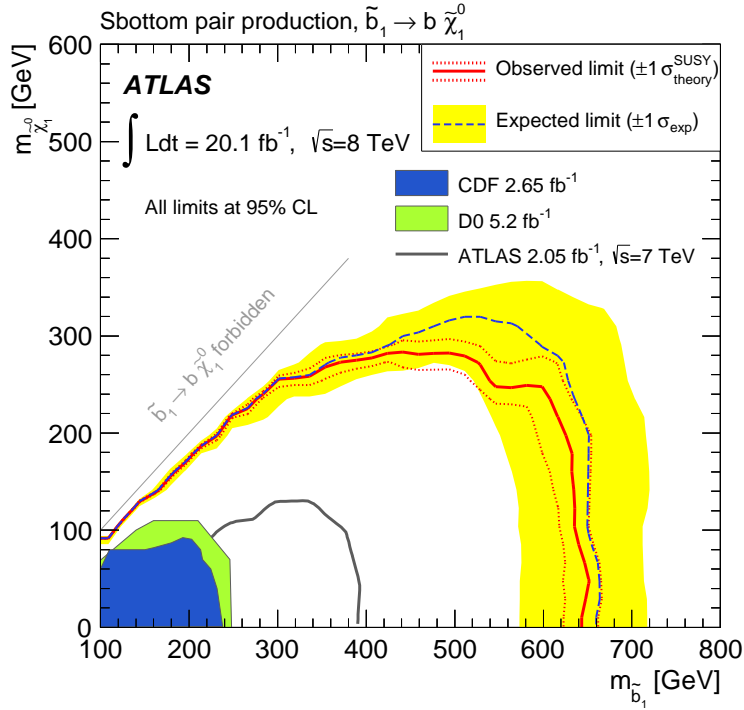}
    \end{center}
\vspace{-0.5cm}
  \caption{Expected and observed exclusion limits at 95\% CL in the $(m_{\bone}, m_{\neut})$ mass plane for the sbottom pair production scenario considered.  
The signal region providing the best expected \CLs exclusion limit is chosen at each point. The dashed (solid) lines show the expected (observed) limits, including all uncertainties except for the theoretical signal cross-section uncertainty (PDF and scale). The bands around the expected limits show the $\pm 1\sigma$ uncertainties. The dotted lines around the observed limits represent the results obtained when moving the nominal signal cross section up or down by the $\pm 1\sigma$ theoretical uncertainty. Previous limits published by ATLAS~\cite{PRL_bjet}, CDF~\cite{Aaltonen:2010dy} and D0~\cite{Abazov:2010wq} are also shown.}
\label{fig:finallimit}
\end{figure}

\begin{figure}[!hpt]
\begin{center}
   \includegraphics[width=0.49\columnwidth]{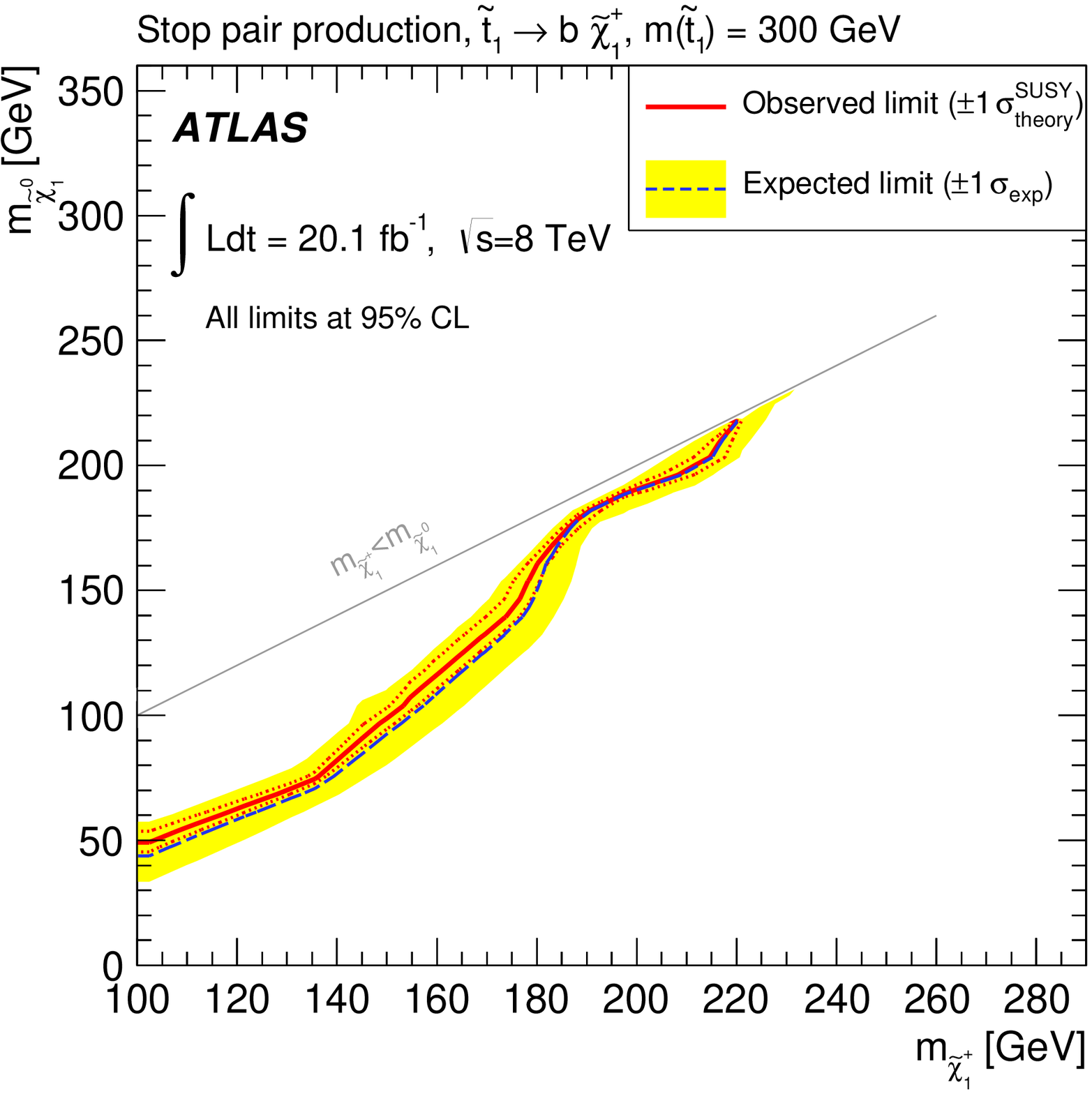}
   \includegraphics[width=0.49\columnwidth]{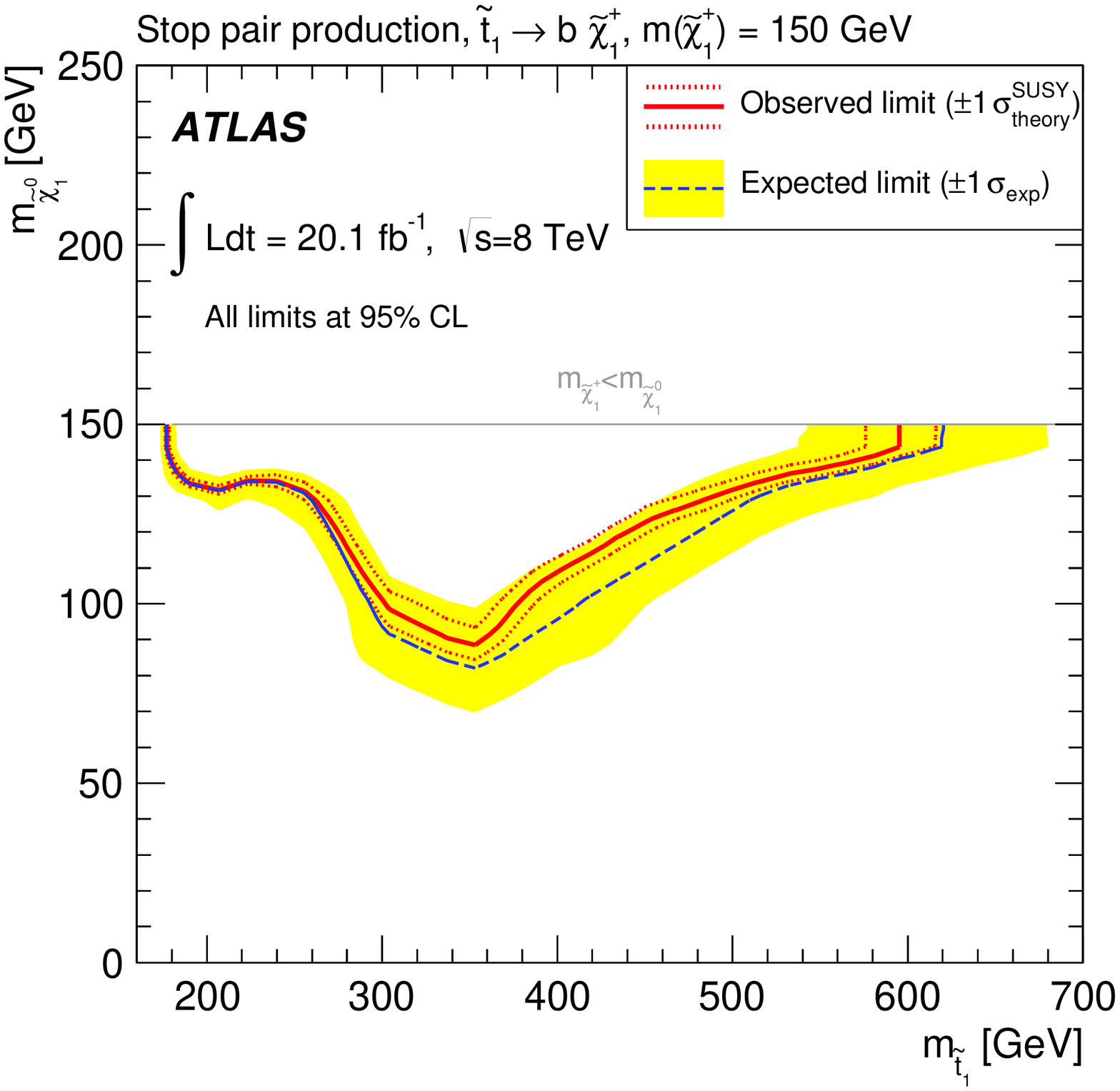}\\
   \includegraphics[width=0.49\columnwidth]{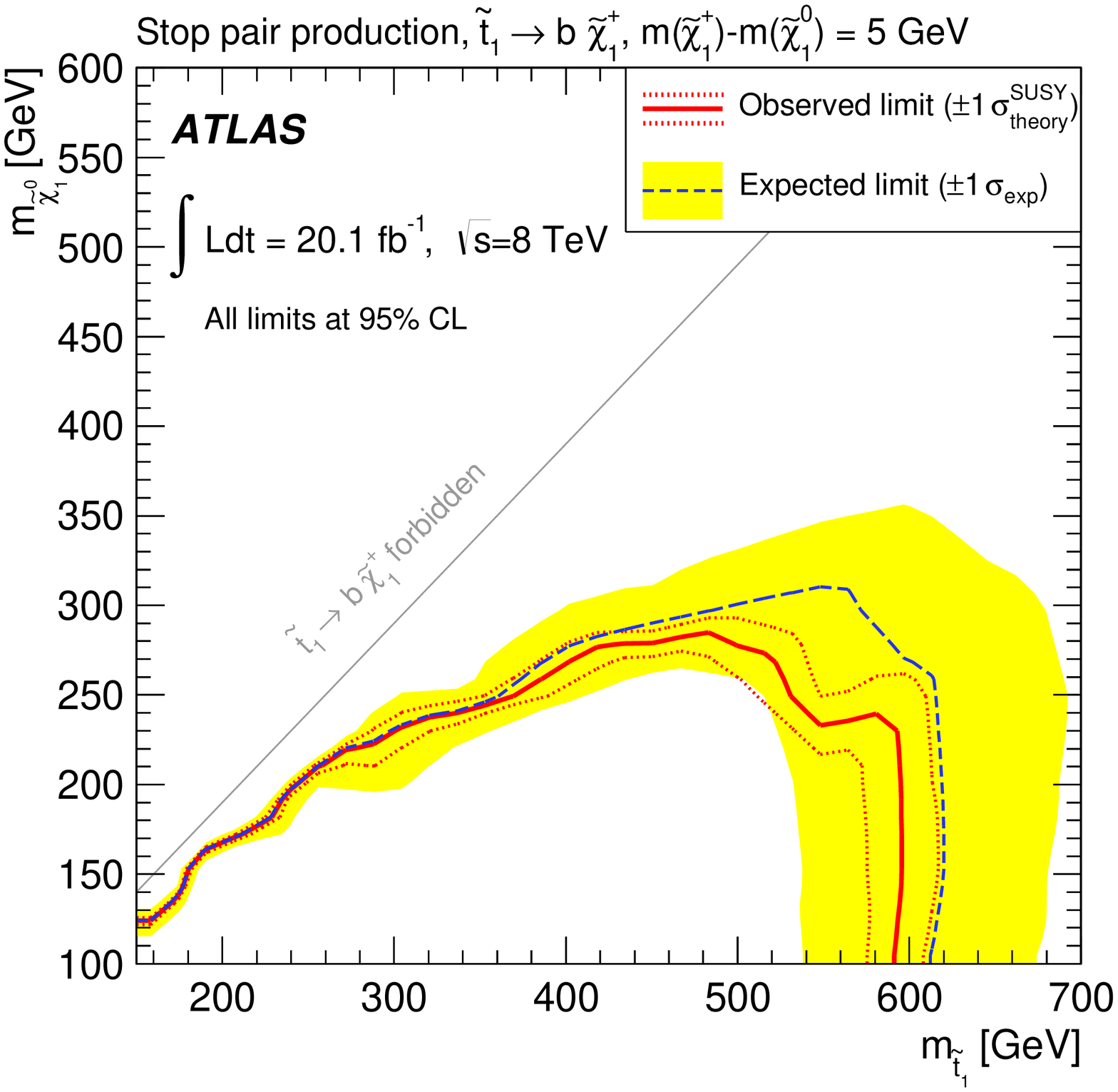}
   \includegraphics[width=0.49\columnwidth]{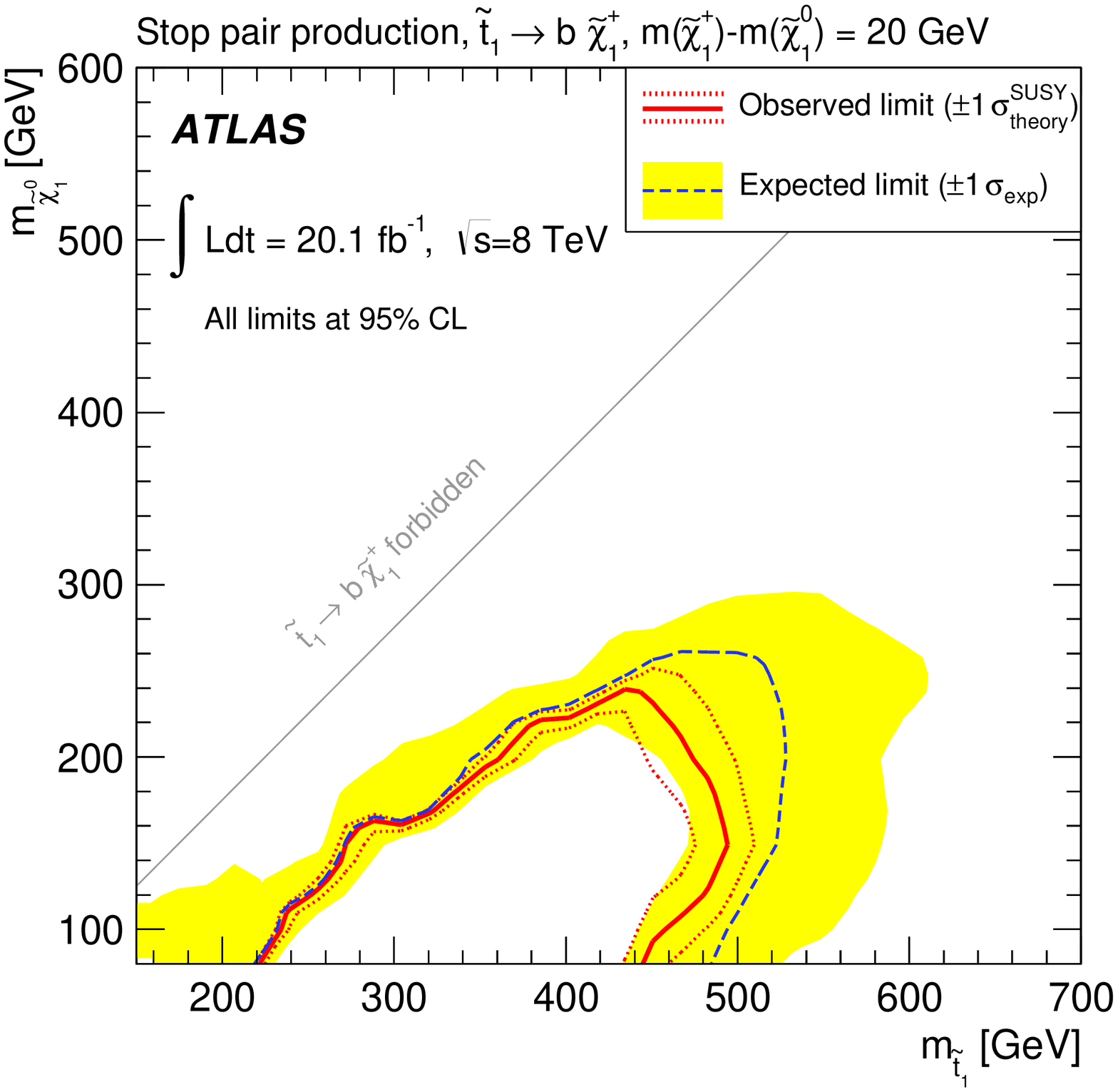}\\
    \end{center}
  \caption{Expected and observed exclusion limits at 95\% CL for the different stop pair production scenarios considered. 
Upper left: $(m_{\chipm}, m_{\neut})$ mass plane with $m_{\tone}=300$~GeV. 
Upper right: $(m_{\tone}, m_{\neut})$ mass plane with $m_{\chipm}=150$~GeV.
Lower left: $(m_{\tone}, m_{\neut})$ mass plane with $m_{\chipm}-m_{\neut}=5$~GeV. Lower right: $(m_{\tone}, m_{\neut})$ mass plane with $m_{\chipm}-m_{\neut}=20$~GeV. The signal region providing the best expected \CLs exclusion limit is chosen at each point. The dashed (solid) lines show the expected (observed) limits, including all uncertainties except for the theoretical signal cross-section uncertainty (PDF and scale). The bands around the expected limits show the $\pm 1\sigma$ uncertainties. The dotted lines around the observed limits represent the results obtained when moving the nominal signal cross section up or down by the $\pm 1\sigma$ theoretical uncertainty. The excluded regions are above (below) the curves for the upper (lower) figures. For the bottom two figures, the lower bound of the vertical axis corresponds to the LEP limit of the lightest chargino mass, 103.5~GeV~\cite{lep2}.}
\label{fig:stoplimit}
\end{figure}

In the case of stop pair production with the stop decaying only into $b \chipm$, the model depends on the masses of the three SUSY particles involved in the decay, $m_{\tone}$, $m_{\chipm}$ and $m_{\neut}$. 
Limits are derived using the same procedure adopted for the sbottom pair production scenario and are presented in 
Figure~\ref{fig:stoplimit} under the additional assumptions that $m_{\tone}$=300~GeV (upper left), 
$m_{\chipm}$=150~GeV (upper right), or for a fixed value of the mass difference $\Delta m = m_{\chipm}-m_{\neut}$ of 5~GeV (lower left) and 20~GeV (lower right).  
Stop masses up to 580~GeV (440~GeV) are excluded for $\Delta m=5$~GeV (20~GeV) and for $m_{\neut}= 100$~GeV. For $\Delta m=5$~GeV (20~GeV), neutralino masses up to 270~GeV (220~GeV) are excluded  for $m_{\tone}= 420$~GeV. In the $\Delta m=20$~GeV, a smaller fraction of the $(m_{\tone}, m_{\neut})$ plane is excluded since this scenario has a lower efficiency given that electrons and muons often have a \pt above the reconstruction threshold.

\section{Conclusions}
The results of a search for third-generation squark
pair production in $pp$ collisions at $\sqrt{s} = 8$~TeV based
on \totallumi\ of ATLAS data are reported.  Events
with large \met~and two \btagged jets are analysed. The results are in agreement with SM
predictions for backgrounds and translate into 95\% CL. 
upper limits on the sbottom (stop) and neutralino masses in a given MSSM 
scenario for which the exclusive decay $\bone \rightarrow b\neut$ ($\tone \rightarrow b \chipm$)
is assumed. For sbottom pairs  decaying exclusively to $b \neut$, sbottom masses up to 620~GeV are excluded at 95\% CL for $m_{\neut} < 150$~GeV. Differences in mass above 50~GeV between the \bone and the \neut are excluded up to sbottom masses of 300~GeV.
 These limits significantly extend previous results.

For stop pairs decaying exclusively into $b \chipm$, stop masses up to 580~GeV (440~GeV) are excluded for $\Delta m=m_{\chipm}-m_{\neut}=$~5~GeV (20~GeV) and for $m_{\neut}= 100$~GeV. For $\Delta m=5$~GeV (20~GeV), neutralino masses up to 270~GeV (220~GeV) are excluded  for $m_{\tilde{t}_1}= 420$~GeV.

\section*{Acknowledgements}


We thank CERN for the very successful operation of the LHC, as well as the
support staff from our institutions without whom ATLAS could not be
operated efficiently.

We acknowledge the support of ANPCyT, Argentina; YerPhI, Armenia; ARC,
Australia; BMWF and FWF, Austria; ANAS, Azerbaijan; SSTC, Belarus; CNPq and FAPESP,
Brazil; NSERC, NRC and CFI, Canada; CERN; CONICYT, Chile; CAS, MOST and NSFC,
China; COLCIENCIAS, Colombia; MSMT CR, MPO CR and VSC CR, Czech Republic;
DNRF, DNSRC and Lundbeck Foundation, Denmark; EPLANET, ERC and NSRF, European Union;
IN2P3-CNRS, CEA-DSM/IRFU, France; GNSF, Georgia; BMBF, DFG, HGF, MPG and AvH
Foundation, Germany; GSRT and NSRF, Greece; ISF, MINERVA, GIF, DIP and Benoziyo Center,
Israel; INFN, Italy; MEXT and JSPS, Japan; CNRST, Morocco; FOM and NWO,
Netherlands; BRF and RCN, Norway; MNiSW, Poland; GRICES and FCT, Portugal; MERYS
(MECTS), Romania; MES of Russia and ROSATOM, Russian Federation; JINR; MSTD,
Serbia; MSSR, Slovakia; ARRS and MIZ\v{S}, Slovenia; DST/NRF, South Africa;
MICINN, Spain; SRC and Wallenberg Foundation, Sweden; SER, SNSF and Cantons of
Bern and Geneva, Switzerland; NSC, Taiwan; TAEK, Turkey; STFC, the Royal
Society and Leverhulme Trust, United Kingdom; DOE and NSF, United States of
America.

The crucial computing support from all WLCG partners is acknowledged
gratefully, in particular from CERN and the ATLAS Tier-1 facilities at
TRIUMF (Canada), NDGF (Denmark, Norway, Sweden), CC-IN2P3 (France),
KIT/GridKA (Germany), INFN-CNAF (Italy), NL-T1 (Netherlands), PIC (Spain),
ASGC (Taiwan), RAL (UK) and BNL (USA) and in the Tier-2 facilities
worldwide.

\clearpage

\bibliographystyle{JHEP}      


\providecommand{\href}[2]{#2}\begingroup\raggedright\endgroup

\onecolumn

\clearpage

\begin{flushleft}
{\Large The ATLAS Collaboration}

\bigskip

G.~Aad$^{\rm 48}$,
T.~Abajyan$^{\rm 21}$,
B.~Abbott$^{\rm 112}$,
J.~Abdallah$^{\rm 12}$,
S.~Abdel~Khalek$^{\rm 116}$,
O.~Abdinov$^{\rm 11}$,
R.~Aben$^{\rm 106}$,
B.~Abi$^{\rm 113}$,
M.~Abolins$^{\rm 89}$,
O.S.~AbouZeid$^{\rm 159}$
% ...
\bigskip
\\
% ...
$^{11}$ Institute of Physics, Azerbaijan Academy of Sciences, Baku, Azerbaijan\\
$^{12}$ Institut de F{\'\i}sica d'Altes Energies and Departament de F{\'\i}sica de la Universitat Aut{\`o}noma de Barcelona, Barcelona, Spain\\
% ...
$^{21}$ Physikalisches Institut, University of Bonn, Bonn, Germany\\
% ...
$^{48}$ Fakult{\"a}t f{\"u}r Mathematik und Physik, Albert-Ludwigs-Universit{\"a}t, Freiburg, Germany\\
% ...
$^{89}$ Department of Physics and Astronomy, Michigan State University, East Lansing MI, United States of America\\
% ...
$^{106}$ Nikhef National Institute for Subatomic Physics and University of Amsterdam, Amsterdam, Netherlands\\
% ...
$^{112}$ Homer L. Dodge Department of Physics and Astronomy, University of Oklahoma, Norman OK, United States of America\\
$^{113}$ Department of Physics, Oklahoma State University, Stillwater OK, United States of America\\
% ...
$^{116}$ LAL, Universit{\'e} Paris-Sud and CNRS/IN2P3, Orsay, France\\
% ...
$^{159}$ Department of Physics, University of Toronto, Toronto ON, Canada\\
% ...
\end{flushleft}


\end{document}